\documentclass[journal]{IEEEtran}

\usepackage[nospace,noadjust]{cite}

\usepackage{amssymb,amsmath}
\usepackage{graphics}
\usepackage{float}
\usepackage{graphicx}
\usepackage{amsmath}
\usepackage{cite}
\usepackage{dsfont}
\usepackage{bm}
\usepackage{upgreek}
\usepackage{arydshln}
\usepackage{enumerate}
\usepackage{amsthm}
\usepackage{graphicx}
\usepackage{threeparttable}
\usepackage{caption}
\usepackage{multirow}
\usepackage{xfrac}
\usepackage{array}
\usepackage{nomencl}
\usepackage{hyphenat}
\usepackage{algorithm,algorithmic}
\usepackage{url}
\usepackage[nice]{nicefrac}
\usepackage[monochrome]{color}
\usepackage{multicol}
\usepackage{setspace}
\usepackage{subcaption}

\newcommand{\be}{\begin{equation}}
\newcommand{\ee}{\end{equation}}

\newcommand{\parder}[2]{ \frac{\partial #1}{\partial #2} }
\newcommand{\norm}[1]{ || #1 ||}
\newcommand{\trace}[1]{ \mathrm{tr}( #1 )}
\newcommand{\mb}[1]{\mathbf{#1}}
\newcommand{\bs}[1]{\boldsymbol{#1}}

\newcommand{\virg}[1]{\textquotedblleft#1\textquotedblright}

\newcommand{\vsigmat}{\mathrm{vecs}(\bs{\Sigma}_0)}

\newcommand{\kronVtinvm}{\mb{V}_1^{-1/2}\otimes\mb{V}_1^{-1/2}}
\newcommand{\kronVtinvmT}{\mb{V}_1^{-T/2}\otimes\mb{V}_1^{-1/2}}

\newcommand{\cvec}[1]{ \mathrm{vec}\left(  #1 \right) }
\newcommand{\vecs}[1]{\mathrm{vecs}(#1)}
\newcommand{\ovec}[1]{\underline{\mathrm{vec}}(#1)}
\newcommand{\ovecs}[1]{\underline{\mathrm{vecs}}(#1)}
\newcommand{\tonde}[1]{\left( #1 \right)  }
\newcommand{\quadre}[1]{\left[  #1 \right]  }
\newcommand{\graffe}[1]{\left\lbrace   #1 \right\rbrace   }

\newtheorem{theorem}{Theorem}

\newtheorem{lemma}{Lemma}

\newtheorem{proposition}{Proposition}

\begin{document}
	
\title{Robust Semiparametric Efficient Estimators in Elliptical Distributions}

\author{Stefano~Fortunati,~\IEEEmembership{Member,~IEEE,}
	Alexandre~Renaux,~\IEEEmembership{Member,~IEEE,}
	Fr\'{e}d\'{e}ric~Pascal,~\IEEEmembership{Senior Member,~IEEE}
	\thanks{S.~Fortunati, A.~Renaux, F.~Pascal are with Universit\'{e} Paris-Saclay, CNRS, CentraleSupel\'{e}c, Laboratoire des signaux et syst\`{e}mes, 91190, Gif-sur-Yvette, France. 
		(e-mails: stefano.fortunati, frederic.pascal@centralesupelec.fr, alexandre.renaux@u-psud.fr).}% <-this % stops a space
	\thanks{The work of S. Fortunati, A. Renaux and F. Pascal has been partially supported by DGA under grant ANR-17-ASTR-0015.}}

\maketitle

\begin{abstract}
Covariance matrices play a major role in statistics, signal processing and machine learning applications. This paper focuses on the \textit{semiparametric} covariance/scatter matrix estimation problem in elliptical distributions. The class of elliptical distributions can be seen as a semiparametric model where the finite-dimensional vector of interest is given by the location vector and by the (vectorized) covariance/scatter matrix, while the density generator represents an infinite-dimensional nuisance function. The main aim of this work is then to provide possible estimators of the finite-dimensional parameter vector able to reconcile the two dichotomic concepts of \textit{robustness} and (semiparametric) \textit{efficiency}. An $R$-estimator satisfying these requirements has been recently proposed by Hallin, Oja and Paindaveine for real-valued elliptical data by exploiting the Le Cam's theory of \textit{one-step efficient estimators} and the \textit{rank-based statistics}. In this paper, we firstly recall the building blocks underlying the derivation of such real-valued $R$-estimator, then its extension to complex-valued data is proposed. Moreover, through numerical simulations, its estimation performance and robustness to outliers are investigated in a finite-sample regime.    
\end{abstract}


\section{Introduction}
\label{Introduction}
Semiparametric inference is the branch of theoretical and applied statistics dealing with point estimation or hypothesis testing in semiparametric model. In short, a semiparametric model is a family of probability density functions (pdfs) parameterized by a finite-dimensional parameter vector of interest, say $\bs{\phi} \in \Omega \subseteq \mathbb{R}^q$ (or $\mathbb{C}^q$), and by an infinite-dimensional parameter, say $g \in \mathcal{G}$, where $\mathcal{G}$ is a suitable set of functions \cite{BKRW}. In the vast majority of applications where semiparametric models are used, the infinite-dimensional parameter $g$ plays the role of a \textit{nuisance} function. 

Despite of their generality and practical relevance, the use of semiparametric models in Signal Processing (SP) applications is still limited to very few cases. To name some examples, we refer to \cite{Amari} for a semiparametric approach to blind source separation, to \cite{Zoubir_semi} for robust non-linear regression and to \cite{PASCAL_el} for empirical likelihood methods applied to covariance estimation. More recently, in \cite{For_SCRB,For_SCRB_complex}, the class of the Real and Complex Elliptically Symmetric (RES and CES) distributions \cite{Esa} has been revised from a semiparametric standpoint (see also \cite{Bickel_paper,Hallin_P_Annals,Hallin_Annals_Stat_2,Hallin_P_2006,PAINDAVEINE} in the statistical literature). %The real and complex ES distributions have been successfully used as a non-Gaussian statistical model able to describe the heavy-tailed behavior of noisy data in different applications such as radar and sonar processing, communications and image processing (see e.g. \cite{Sang, zozor, RichPhD, Esa, book_zoubir}). 
The family of Elliptically Symmetric (ES) distributions is in fact a typical example of semiparametric model where the finite-dimensional parameter vector of interest is given by the location vector $\bs{\mu}$ and by the (vectorized version of) the covariance/scatter matrix $\bs{\Sigma}$, while the density generator $g$ can be considered as a nuisance function. In particular, in \cite{For_SCRB} the RES class has been framed in the context of \textit{semiparametric group models}, then a Semiparametric Cram\'{e}r-Rao Bound (SCRB) for the joint estimation of $\bs{\mu}$ and $\bs{\Sigma}$ in the presence of the nuisance density generator $g$ has been derived. The second work \cite{For_SCRB_complex} extended the previously obtained SCRB to semiparametric estimation of complex parameters in CES distributed data. A semiparametric version of the celebrated Slepian-Bangs formula has been also proposed. However, the following fundamental question has not been addressed in \cite{For_SCRB,For_SCRB_complex} which were focused on lower bounds: \textit{is it possible to derive a robust and semiparametric efficient estimator of the covariance/scatter matrix $\bs{\Sigma}$ of a set of ES distributed observations?} \textcolor{blue}{As we will see ahead, a first positive answer to this question has been provided in \cite{Hallin_Annals_Stat_2} for the RES case while its extension to CES distributions will be given in this paper.}

%The problem of estimating the covariance/scatter matrix of a set of acquired observations is ubiquitous in applications: from signal and array processing, machine learning and data analysis, to geophysics and econometrics. As a consequence, a lot of practitioners may benefit from having such a robust and semiparametric efficient estimator in their toolbox. 

To start, let us take a closer look to the two main features that this estimator should have. Firstly, it should be \textit{semiparametric efficient}, at least asymptotically. In other words, we require that the error covariance matrix of such estimator should be equal to the SCRB given in \cite{For_SCRB,For_SCRB_complex} as the number of observations goes to infinity. The second desirable feature is the \textit{distributional robustness}. As said before, a semiparametric model allows for the presence of a nuisance function that, in the case of ES distributed observations, is the unknown density generator $g$ characterizing the shape of their actual distribution. So, a distributionally robust estimator is basically an estimator of $\bs{\Sigma}$ whose statistical properties do not rely on $g \in \mathcal{G}$, and consequently on the actual ES distribution of the data. It is worth to underline that, even if robust estimators of covariance matrices are already available in the statistics and SP literature (\cite{Tyler1,Gini1,Pascal1,Pascal2,Esa,draskovic}, \cite[Ch. 4]{book_zoubir} and references therein), they fail to be semiparametric efficient as shown in \cite{For_SCRB,For_SCRB_complex}.

A good candidate for the estimator that we are looking for is the one proposed by Hallin, Oja and Paindaveine in their seminal paper \cite{Hallin_Annals_Stat_2}. Building upon their previous work \cite{Hallin_P_Annals}, in \cite{Hallin_Annals_Stat_2} the Authors propose an estimator of the constrained, real-valued scatter matrix $\bs{\Sigma}$ in RES distributed data that meets the two requirements of \textit{nearly} semiparametric efficiency and distributional robustness. To achieve the semiparametric efficiency, the Le Cam's theory of \textit{one-step efficient estimators} \cite{LeCam60}, \cite[Ch. 6]{LeCam} has been exploited. In fact, as showed by Le Cam, it is possible to derive asymptotically efficient estimators that, unlike the Maximum Likelihood (ML) one, do not search for the maxima of the log-likelihood function. This is of great importance in practical applications, where the ML estimator can present computational difficulties in the resulting optimization problem or even existence/uniqueness issues \cite[Ch. 6]{Lehm98}. The second requirement of distributional robustness has been addressed in \cite{Hallin_Annals_Stat_2} using a rank-based approach \cite{hajek1968}, \cite[Ch. 13]{vaart_1998}. Originally developed in the context of order statistics, rank-based methods has been used in robust statistics to derive distributionally robust estimators and tests that are usually referred to as $R$-estimators and $R$-tests \cite[Ch. 3]{huber_book}. 
          
\textcolor{blue}{After a semiparametric formalization of the shape matrix estimation problem given in Section \ref{preliminaries}, the subsequent Section \ref{R_est_sec} provides a review of the methodology used in \cite{Hallin_Annals_Stat_2} to derive a semiparametric efficient $R$-estimator of the constrained, real-valued, scatter matrix $\bs{\Sigma}$ in RES distributed data. This first part has the twofold goal of \textit{i}) introducing two statistical procedures (i.e.\ semiparametric one-step estimators and rank-based robustification) that are not yet widespread among the SP community and then \textit{ii}) showing how they can be applied to derive original estimators of scatter matrices. To this end, additional in-depth supporting material will be provided separately from the main body of the paper. In addition, the code containing our Matlab and Python implementation of both real- and complex-valued $R$-estimator can be found at \cite{Code_R}. Section \ref{complex_ext} focuses on the extension of the previously derived outcomes to the complex-valued parameter case with Complex ES distributed data. In Section \ref{numerical} the Mean Squared Error (MSE) performance and the robustness properties of the proposed semiparametric efficient $R$-estimator will be investigated through numerical simulations in a \virg{finite-sample} regime. The theoretical analysis, in fact, can only provide us with asymptotic guarantees on the good behavior of an estimator but, since in practice the number of available observation is always finite, a \virg{finite-sample} performance characterization is necessary as well. To this end, the error covariance matrix of the proposed $R$-estimator (evaluated using independent Monte Carlo runs) will be compared with the SCRB in \cite{For_SCRB, For_SCRB_complex} in different scenarios. The second feature that is going to be assessed in Section \ref{numerical} is the robustness to the presence of \textit{outliers} in the observations. In the present context, an outlier can be represented by an observation vector whose distribution does not belong to the ES family.} 
%By means of numerical simulations, we show that the robustness to outliers of the proposed $R$-estimator is comparable to the one of other celebrated robust $M$-estimators of scatter matrices as the Tyler's one \cite{Tyler1,Esa}. A more rigorous characterization of the robustness to outliers based on e.g. the influence function and the breakdown point \cite{Hampel,huber_book}, is left to future work.

%In the rest of the paper, we assume that the reader is already familiar with the basic concepts of the semiparametric theory, and in particular with the definition of \textit{efficient score function} and \textit{efficient semiparametric Fisher Information Matrix} (SFIM) that can be found in the relevant statistical literature (e.g. in the monographs \cite{BKRW, Tsiatis}) or in \cite{For_EUSIPCO,For_SCRB,For_SCRB_complex} where they are introduced in a more familiar way for the SP community.

\textit{Algebraic notation}: Throughout this paper, italics indicates scalar quantities ($a$), lower case and upper case boldface indicate column vectors ($\mathbf{a}$) and matrices ($\mathbf{A}$), respectively. Each entry of a matrix $\mb{A}$ is indicated as $a_{ij}\triangleq [\mb{A}]_{i,j}$. $\mb{I}_N$ defines the $N \times N$ identity matrix. The superscripts $*$, $\top$ and $\mathsf{H}$ indicate the complex conjugation, the transpose and the Hermitian operators respectively, then ${{\mathbf{A}}^\mathsf{H}} = {({{\mathbf{A}}^ * })^\top}$. Moreover, $\mb{A}^{- \top} \triangleq (\mb{A}^{-1})^\top = (\mb{A}^\top)^{-1}$, $\mb{A}^{-*} \triangleq (\mb{A}^{-1})^* = (\mb{A}^*)^{-1}$ and $\mb{A}^{-\mathsf{H}} \triangleq (\mb{A}^{-1})^\mathsf{H} = (\mb{A}^\mathsf{H})^{-1}$. The Euclidean norm of a vector $\mb{a}$ is indicated as $\norm{\mb{a}}$. The determinant and the Frobenius norm of a matrix $\mb{A}$ are indicated as $|\mb{A}|$ and $\norm{\mb{A}}_F$, respectively. The symbol $\mathrm{vec}$ indicates the standard vectorization operator that maps column-wise the entry of an $N \times N$ matrix $\mathbf{A}$ in an $N^2$-dimensional column vector $\cvec{\mb{A}}$. The operator $\ovec{\mb{A}}$ defines the $N^2-1$-dimensional vector obtained from $\cvec{\mb{A}}$ by deleting its first element, i.e. $\cvec{\mb{A}} \triangleq [a_{11},\ovec{\mb{A}}^\top]^\top$. A matrix $\mb{A}$ whose first top-left entry is constrained to be equal to 1, i.e. $a_{11} \triangleq 1$, is indicated as $\mb{A}_1$. 
%The matrix
%\be
%\label{mat_proj_I}
%\Pi^{\perp}_{\cvec{\mb{I}_N}}=\mb{I}_{N^2} - N^{-1}\mathrm{vec}(\mb{I}_N)\mathrm{vec}(\mb{I}_N)^\top.
%\ee
%is the orthogonal projection matrix on the orthogonal complement of $\mathrm{span}(\mathrm{vec}(\mb{I}_N))$.

For any $N \times N$ symmetric matrix $\mathbf{A}$: 
\begin{itemize}
	\item $\vecs{\mb{A}}$ indicates the $N(N+1)/2$-dimensional vector of the entries of the lower (or upper) sub-matrix of $\mathbf{A}$.
	\item According to the notation previously introduced, $\vecs{\mb{A}} \triangleq [a_{11},\ovecs{\mb{A}}^\top]^\top$.
	\item If $a_{11} = 0$, then $\mb{M}_N$ is the $N(N+1)/2-1 \times N^2$ matrix such that (s.t.) $\mb{M}_N^\top\ovecs{\mb{A}} = \cvec{\mb{A}}$. Note that $\mb{M}_N^\top$ can be obtained from the \textit{duplication matrix} $\mb{D}_N$ \cite{Magnus1, Magnus2} by removing its first column.
	%\item Lastly, $\mb{N}_N$ indicates the $N(N+1)/2-1 \times N^2$ matrix s.t. $\mb{N}_N^\top\cvec{\mb{A}} = \ovecs{\mb{A}}$. Again, $\mb{N}_N$ can be obtained from the \textit{elimination matrix} $\mb{E}_N$ \cite{Magnus1, Magnus2} by removing its first column.  
\end{itemize}

\textit{Statistical notation}: Let $x_l$ be a sequence of random variables in the same probability space. We write:
\begin{itemize}
	\item $x_l = o_P(1)$ if $\lim_{l\rightarrow \infty}\mathrm{Pr}\graffe{|x_l|\geq\epsilon}=0,\forall \epsilon>0$ (\textit{convergence in probability to 0}),
 	\item $x_l = O_P(1)$ if for any $\epsilon > 0$, there exists a finite $M>0$ and a finite $L>0$, s.t. $\mathrm{Pr}\graffe{|x_l|>M}<\epsilon,\forall l>L$ (\textit{stochastic boundedness}).
\end{itemize}
The cumulative distribution function (cdf) and the related probability density function (pdf) of a random variable $x$ or a random vector $\mb{x}$ are indicated as $P_X$ and $p_X$, respectively. For random variables and vectors, $\overset{d}{=}$ stands for \virg{has the same distribution as}. The symbol $\underset{L \rightarrow \infty}{\sim}$ indicates the convergence in distribution. According to the notation introduced in \cite{For_EUSIPCO, For_SCRB, For_SCRB_complex}, we indicate the \textit{true} pdf as $p_0(\mb{x})\triangleq p_X(\mb{x}|\bs{\phi}_0, g_0)$, where $\bs{\phi}_0$ and $g_0$ indicate the true parameter vector to be estimated and the true nuisance function, respectively. We define as $E_{\bs{\phi}, g}\{f(\mb{x})\} = \int f(\mb{x})p_X(\mb{x}|\bs{\phi}, g) d\mb{x}$ the expectation operator of a \textit{measurable} function $f$ of a random vector $\mb{x}$. Moreover, we simply indicate as $E_0\{\cdot\}$ the expectation with respect to (w.r.t.) the true pdf $p_0(\mb{x})$. The superscript $\star$ indicates a $\sqrt{L}$-consistent, \textit{preliminary}, estimator $\hat{\bs{\phi}}^\star$ of $\bs{\phi}_0$, s.t. $\sqrt{L} \tonde{\bs{\phi}^\star - \bs{\phi}_0} = O_{P}(1)$. \textcolor{blue}{The dependence of $\mb{x}$ of a function $f(\mb{x})$ is often dropped for notation simplicity: $f \equiv  f(\mb{x})$.} 
 
\section{The semiparametric shape matrix estimation}
\label{preliminaries}
Let $\{\mb{x}_l\}_{l=1}^L$ be a set of $N$-dimensional, real-valued, \textit{independent} and \textit{identically distributed} (i.i.d.) observation vectors. Each observation is assumed to be sampled from a real elliptical pdf \cite{CAMBANIS1981,RES_Fang,Esa} of the form:
\be
\label{RES_pdf}
p_X(\mb{x}_l|\bs{\mu},\bs{\Sigma},g)=2^{-N/2}|\bs{\Sigma}|^{-1/2} g \left((\mb{x}_l-\bs{\mu})^\top\bs{\Sigma}^{-1}(\mb{x}_l-\bs{\mu}) \right),
\ee
where $\bs{\mu} \in \mathbb{R}^N$ is a location vector, $\bs{\Sigma} \in \mathcal{M}_N^\mathbb{R}$ is a $N \times N$ \textit{scatter} matrix in the set $\mathcal{M}_N^\mathbb{R}$ of the symmetric, positive definite, real matrices. The function $g \in \mathcal{G}$ is the \textit{density generator}, an infinite-dimensional parameter that characterizes the specific distribution in the RES family. In order to guarantee the integrability of the pdf in \eqref{RES_pdf},    
the set of all the possible density generators is defined as $\mathcal{G} = \graffe{ g: \mathbb{R}^{+} \rightarrow \mathbb{R}^{+} \left|   \int_{0}^{\infty}t^{N/2-1}g(t)dt < \infty, \int p_Xd\mb{x} = 1 \right. }$\cite{CAMBANIS1981}.
%\be
%\label{set_G}
%\mathcal{G} = \graffe{ g: \mathbb{R}_0^{+} \rightarrow \mathbb{R}^{+} \left|   \int_{0}^{\infty}t^{N/2-1}g(t)dt < \infty \right. }.
%\ee
Each random vector whose pdf is given by \eqref{RES_pdf}, say $\mb{x} \sim RES_N(\bs{\mu},\bs{\Sigma},g)$, admits the following \textit{stochastic representation} \cite{CAMBANIS1981,Esa}:
\be
\label{SRT_dec}
\mb{x} \overset{d}{=} \bs{\mu} + \mathcal{R}\bs{\Sigma}^{1/2}\mb{u},
\ee 
where $\mb{u} \sim \mathcal{U}(\mathbb{R}S^{N-1})$ is uniformly distributed on the unit $(N-1)$-sphere \textcolor{blue}{$\mathbb{R}S^{N-1} \triangleq \{\mb{u}\in \mathbb{R}^N|\norm{\mb{u}}=1\}$}, $\mathcal{R} \triangleq \sqrt{\mathcal{Q}}$ is called \textit{modular variate} while $\mathcal{Q}$, usually referred to as \textit{2nd-order modular variate}, is such that (s.t.)
\be
\label{Q_RES}
\mathcal{Q}\overset{d}{=} (\mb{x}_l-\bs{\mu})^\top\bs{\Sigma}^{-1}(\mb{x}_l-\bs{\mu})\triangleq Q_l,\forall l.
\ee
Moreover, $\mathcal{Q}$ has pdf given by:
\be
\label{SS_Q_pdf}
p_{\mathcal{Q}}(q) = (\pi/2)^{N/2}\Gamma(N/2)^{-1} q^{N/2-1} g (q),
\ee
where $\Gamma(\cdot)$ stands for the Gamma function.
%The two positive, real-valued, random variables $\mathcal{R}$ and $\mathcal{Q}$ admit pdfs given by:
%\be
%\label{SS_Q_pdf}
%p_{\mathcal{Q}}(q) = s_N 2^{-N/2-1} q^{N/2-1} g \left(q \right), 
%\ee
%\be
%\label{SS_R_pdf}
%p_{\mathcal{R}}(r) = s_N 2^{-N/2} r^{N-1} g \left(r^2 \right).
%\ee
%where $s_N \triangleq 2 \pi^{N/2}/\Gamma(N/2)$.
 
The expression of the elliptical pdf in \eqref{RES_pdf} and the stochastic representation in \eqref{SRT_dec} are not uniquely defined due to the well-know scale ambiguity between the scatter matrix $\bs{\Sigma}$ and the density generator $g$. Specifically, from \eqref{RES_pdf}, it is immediate to verify that $RES_{N}(\bs{\mu},\bs{\Sigma},g(t))\equiv RES_{N}(\bs{\mu},c\bs{\Sigma},g(t/c)), \forall c>0$. In an equivalent way, from \eqref{SRT_dec}, we have that $\mb{x} \overset{d}{=} \bs{\mu} + \mathcal{R}\bs{\Sigma}^{1/2}\mb{u}\overset{d}{=} \bs{\mu} + (c^{-1}\mathcal{R})(c\bs{\Sigma}^{1/2})\mb{u}, \forall c>0$. This readily implies that $\bs{\Sigma}$ is identifiable only up to a scale factor and consequently only a \textit{scaled} version of $\bs{\Sigma}$ can be estimated. To avoid this identifiability problem, following \cite{Hallin_P_2006,PAINDAVEINE,Esa}, let us define the symmetric and positive definite \textit{shape} matrix $\mb{V}$ as:
\be
\label{shape}
\mb{V} \triangleq \bs{\Sigma}/s(\bs{\Sigma}),
\ee
where $s : \mathcal{M}_N^\mathbb{R} \rightarrow \mathbb{R}^{+}$ is a scalar functional on $\mathcal{M}_N^\mathbb{R}$ satisfying the following assumptions \cite{Hallin_P_2006,PAINDAVEINE}:
\begin{itemize}
	\item[A1] Homogeneity: $s(c \cdot \bs{\Sigma}) = c \cdot s(\bs{\Sigma}), \forall c>0$,
	\item[A2] Differentiability over $\mathcal{M}_N^\mathbb{R}$ with $\parder{s(\bs{\Sigma})}{[\bs{\Sigma}]_{1,1}} \ne 0$,
	\item[A3] $s(\mb{I}_N) = 1$.
\end{itemize} 
Typical examples of this class of scale functional are $s(\bs{\Sigma}) = [\bs{\Sigma}]_{1,1}$, $s(\bs{\Sigma}) = \trace{\bs{\Sigma}}/N$ and $s(\bs{\Sigma}) = |\bs{\Sigma}|^{1/N}$. Each scale functional $s$ correspond to a differentiable constraint on the shape matrix $\mb{V}$. As an example, the constraints induced by the three above-mentioned scale functionals are $v_{11} = 1$, $\trace{\mb{V}} = N$ and $|\mb{V}|^{1/N}=1$. It is easy to verify that, under A1, A2 and A3, the first top-left entry of $\mb{V}$, i.e. $v_{11}$, can always be expressed as function of the other entries. 
%Again, if we choose as scale functional $s(\bs{\Sigma}) = [\bs{\Sigma}]_{1,1}$, $v_{11}$ is trivially given by $v_{11} = 1$. For $s(\bs{\Sigma}) = \trace{\bs{\Sigma}}/N$, we have that $v_{11} = N-\sum\nolimits_{n=2}^Nv_{n,n}$. Lastly, for $s(\bs{\Sigma}) = |\bs{\Sigma}|^{1/N}$ it can be easily shown that, using the Laplace's expansion of the determinant along the first row of $\mb{V}$, $v_{11}$ can be recovered as $v_{11} = \frac{1}{C_{1,1}}\tonde{1-\sum\nolimits_{n=2}^Nv_{1,n}C_{1,n}}$ where $C_{i,j}$ indicates the cofactor of $v_{ij}$. 
This consideration, along with the fact that $\mb{V}$ is symmetric by definition, suggests us that, to avoid the identifiability problem, in the semiparamtric estimation problem, we just need to consider the vector $\ovecs{\mb{V}}$ as unknown. Moreover, as discussed in \cite{Hallin_P_2006} and verified here in Sec. \ref{numerical}, the optimality properties of the proposed semiparametric estimator of the shape matrix do not depend on the particular scale functional. Consequently, in order to avoid tedious matrix calculation that may confuse the derivation of the algorithm, we choose the simple scale functional $s(\bs{\Sigma}) = [\bs{\Sigma}]_{1,1}$, i.e. the one that constrains the shape matrix $\mb{V}$ to have its first top-left entry equal to 1. In the rest of the paper, a generic shape matrix satisfying this constraint is indicated as $\mb{V}_1$ according with the notation previously introduced.  

Having said that, we can formally state the semiparametric estimation problem that we are going to analyze in the following sections. Let $\Omega \subseteq \mathbb{R}^q$ be a parameter space of dimension $q = N(N+3)/2-1$ ($=N+N(N+1)/2-1$ where the \virg{$-1$} term is due to the 1-dimensional scale constraint). Each element of $\Omega$ is a vector $\bs{\phi}$ of the form:
\be
\label{def_phi}
\bs{\phi} \triangleq \tonde{\bs{\mu}^\top,\ovecs{\mb{V}_1}^\top}^\top,
\ee
where $\bs{\mu} \in \mathbb{R}^N$ and $\mb{V}_1 \in \mathcal{M}_N^\mathbb{R}$. Let us define the RES semiparametric model as the the following set of (uniquely defined) pdfs:
\be
\label{RES_semi_par_model}
\begin{split}
	\mathcal{P}_{\bs{\phi},g} &= \left\lbrace  p_X | p_X(\mb{x}|\bs{\phi},g) = 2^{-N/2}|\mb{V}_1|^{-1/2} \times \right.  \\ 
	&\left.  g \left((\mb{x}_l-\bs{\mu})^\top\mb{V}_1^{-1}(\mb{x}_l-\bs{\mu}) \right);  \bs{\phi} \in \Omega, g \in \mathcal{G} \right\rbrace .
\end{split}
\ee 
The semiparametric estimation problem that we want to address is then to find a robust and semiparametric efficient estimator of a true parameter vector $\bs{\phi}_0 \in \Omega$ in the presence of a nuisance function $g_0 \in \mathcal{G}$.

\section{\textcolor{blue}{An $R$-estimator for shape matrices in RES data}}\label{R_est_sec}
The aim of this section is to trace the procedure adopted in \cite{Hallin_Annals_Stat_2} to derive the $R$-estimator of real-valued scatter matrices in RES data. In particular, the concepts of Le Cam's one-step estimators and ranks-based robustification will be firstly introduced and their application to the particular semiparametric estimation problem at hand discussed. Finally, a ready-to-use expression of the resulting $R$-estimator is provided, while the related Matlab and Python implementation is given in \cite{Code_R}.

\subsection{Semiparametric efficient one-step estimators} 
The main ingredient for the derivation of a one-step estimator for the parametric part (location vector and scatter matrix) of the semiparametric RES model $\mathcal{P}_{\bs{\phi},g}$ in \eqref{RES_semi_par_model} is the notion of \textit{efficient} score vector. Specifically, the efficient score vector $\bar{\mb{s}}_{\bs{\phi},g_0}$ for the estimation of $\bs{\phi} \in \Omega$ in the presence of a nuisance density generator $g_0 \in \mathcal{G}$ is given by \cite{For_EUSIPCO}, \cite[Th. IV.1]{For_SCRB}:
\be\label{eff_score_vect}
\bar{\mb{s}}_{\bs{\phi},g_0}(\mb{x}_l) \equiv \bar{\mb{s}}_{\bs{\phi},g_0} \triangleq \mb{s}_{\bs{\phi},g_0} - \Pi(\mb{s}_{\bs{\phi},g_0}|\mathcal{T}_{g_0}),
\ee
where $\mb{s}_{\bs{\phi},g_0}(\mb{x}_l)$ is the usual score vector defined as:
\be
\label{score_vect}
\mb{s}_{\bs{\phi},g_0}(\mb{x}_l)=\nabla_{\bs{\phi}}\ln p_X(\mb{x}_l|\bs{\phi},g_0) = \left( \begin{array}{c}
	\mb{s}_{\bs{\mu},g_0}(\mb{x}_l)\\
	\mb{s}_{\ovecs{\mb{V}_1},g_0}(\mb{x}_l)\end{array}\right),
\ee   
and $\Pi(\mb{s}_{\bs{\phi},g_0}|\mathcal{T}_{g_0})$ is the orthogonal projection of the score vector $\mb{s}_{\bs{\phi},g_0}$ in \eqref{score_vect} on the semiparametric nuisance tangent space $\mathcal{T}_{g_0}$ \cite{Tutorial,For_SCRB}. Then, the semiparametric counterpart of the Fisher Information Matrix (FIM) is the \textit{efficient} semiparametric FIM (SFIM) \cite{For_EUSIPCO},\cite[Th. IV.1]{For_SCRB}:
\be\label{S_E_FIM}
\bar{\mb{I}}(\bs{\phi}|g_0) \triangleq E_{\bs{\phi},g_0}\{\bar{\mb{s}}_{\bs{\phi},g_0}(\mb{x})\bar{\mb{s}}_{\bs{\phi},g_0}(\mb{x})^\top\}.
\ee
Finally, we introduce the \textit{efficient central sequence} as:
\be\label{cent_seq_sem}
\overline{\bs{\Delta}}_{\bs{\phi},g_0}(\mb{x}_1,\ldots,\mb{x}_L) \equiv \overline{\bs{\Delta}}_{\bs{\phi},g_0} \triangleq L^{-1/2}\sum\nolimits_{l=1}^{L}\bar{\mb{s}}_{\bs{\phi},g_0}(\mb{x}_l).
\ee
Note that the previous three quantities depend on the true, and generally unknown, density generator $g_0$.

The next Theorem provides us with the expression of the one-step estimator of $\bs{\phi}$ together with its asymptotic properties.
\begin{theorem}
	\label{theo_one_step_semipar}
	Let $\{\mb{x}_l\}_{l=1}^L$ be a set of i.i.d. observations sampled from a RES distribution whose pdf $p_0(\mb{x}) \in \mathcal{P}_{\bs{\phi},g}$ in \eqref{RES_semi_par_model}. Let $\hat{\bs{\phi}}^\star$ be any preliminary $\sqrt{L}$-consistent estimator of the true parameter vector $\bs{\phi}_0\triangleq \tonde{\bs{\mu}_0^\top,\ovecs{\mb{V}_{1,0}}^\top}^\top$. Then, the semiparametric one-step estimator
	\be
	\label{one_step_par_semi}
	\hat{\bs{\phi}}_s = \hat{\bs{\phi}}^\star + L^{-1/2}\bar{\mb{I}}(\hat{\bs{\phi}}^\star|g_0)^{-1}\overline{\bs{\Delta}}_{\hat{\bs{\phi}}^\star,g_0},
	\ee
	has the following properties:
	\begin{itemize}
		\item[PS1] $\sqrt{L}$-consistency
		\be
		\sqrt{L} \tonde{\hat{\bs{\phi}}_s - \bs{\phi}_0} = O_{P}(1),
		\ee
		\item[PS2] Asymptotic normality and efficiency
		\be
		\sqrt{L}\tonde{\hat{\bs{\phi}}_s - \bs{\phi}_0} \underset{L \rightarrow \infty}{\sim} \mathcal{N}(\mb{0},\bar{\mb{I}}(\bs{\phi}_0|g_0)^{-1}),
		\ee
		where $\bar{\mb{I}}(\bs{\phi}_0|g_0)^{-1} = \mathrm{CSCRB}(\bs{\mu}_0,\mb{V}_{1,0}|g_0)$ and the constrained semiparametric CRB (CSCRB) \cite{For_SCRB} is evaluated for the constraint $[\mb{V}_{1,0}]_{11}=1$.
	\end{itemize}
\end{theorem}
\textit{Remark}: The proof of Theorem \ref{theo_one_step_semipar} is given in \cite{Hallin_Annals_Stat_2} (see the proof of the Proposition 2.1). In addition, we refer the interested reader to our supporting material for a tutorial introduction of the Le Cam's theory underlying it. 

Even if semiparametric efficient, the \virg{clairvoyant} estimator $\hat{\bs{\phi}}_s$ in \eqref{one_step_par_semi} relies on the true, and generally unknown, density generator $g_0$, so it is not useful for practical inference problems. Consequently, a distributionally robust alternative to $\hat{\bs{\phi}}_s$ has to be derived, at the price of a possible loss in efficiency. Before addressing the crucial issue of robustness, we provide a \virg{tangible} expression of the clairvoyant estimator of $\mb{V}_1$ that will be useful ahead.
     
\subsection{Semiparametric clairvoyant estimator of shape matrices}
To construct $\hat{\bs{\phi}}_s$ in \eqref{one_step_par_semi} we need explicit expressions of the efficient score vector $\bar{\mb{s}}_{\bs{\phi},g_0}=(\bar{\mb{s}}_{\bs{\mu},g_0}^\top,\bar{\mb{s}}_{\ovecs{\mb{V}_1},g_0}^\top)^\top$, the efficient SFIM $\bar{\mb{I}}(\bs{\phi}|g_0)$ and a preliminary $\sqrt{L}$-consistent estimators $\hat{\bs{\phi}}^\star$ of $\bs{\phi}_0$. Building upon the results in our previous work \cite{For_SCRB}, $\bar{\mb{s}}_{\bs{\mu}}$ and $\bar{\mb{s}}_{\ovecs{\mb{V}_1}}$ can be expressed as \cite[Eq. (53)]{For_SCRB}:
\be
\label{eff_score_mu}
\bar{\mb{s}}_{\bs{\mu},g_0} = \mb{s}_{\bs{\mu},g_0} =  -2 \sqrt{Q_l} \psi_0(Q_l) \mb{V}_1^{-1/2} \mb{u}_l,
\ee
\be
\label{eff_score_V}
\bar{\mb{s}}_{\ovecs{\mb{V}_1},g_0} = -Q_l\psi_0(Q_l)\mb{K}_{\mb{V}_1}  \mathrm{vec}(\mb{u}_l\mb{u}_l^\top),
%\begin{split}
%	\mb{s}_{\ovecs{\mb{V}_1}}(\mb{x}_l) &= - Q_l\psi_0(Q_l) \mb{M}_N \left( \kronVtinvm \mathrm{vec}(\mb{u}_l\mb{u}_l^\top) - \frac{1}{N}\vcVtinv \right) \\
%	& = -Q_l\psi_0(Q_l)\mb{K}_{\mb{V}_1}  \mathrm{vec}(\mb{u}_l\mb{u}_l^\top),\\
%\end{split}
\ee
where $Q_l$ is defined in \eqref{Q_RES} and
\be
\label{K_mat}
\mb{K}_{\mb{V}_1} \triangleq \mb{M}_N \tonde{\kronVtinvm} \Pi^{\perp}_{\cvec{\mb{I}_N}},
\ee
\be
\mb{u}_l \triangleq (Q_l\mb{V}_1)^{-1/2}(\mb{x}_l-\bs{\mu}),
\ee
\be
\label{psi}
\psi_0(t) \triangleq d\ln g_0(t)/dt,
\ee
\be
\label{mat_proj_I}
\Pi^{\perp}_{\cvec{\mb{I}_N}}=\mb{I}_{N^2} - N^{-1}\mathrm{vec}(\mb{I}_N)\mathrm{vec}(\mb{I}_N)^\top,
\ee
where $\mb{M}_N$ is defined in the notation section.
Before moving forward, some comments are in order. As already proved in \cite{For_SCRB}, the efficient score vector $\bar{\mb{s}}_{\bs{\mu},g_0}$ in \eqref{eff_score_mu} of the mean vector is equal to the score vector $\mb{s}_{\bs{\mu},g_0}$, or in other words, $\bar{\mb{s}}_{\bs{\mu},g_0}$ is orthogonal to the nuisance tangent space $\mathcal{T}_{g_0}$. This implies that, knowing or not knowing the true density generator $g_0$ does not have any impact on the asymptotic performance of an estimator of $\bs{\mu}$. The expression of the efficient score vector for the shape matrix in Eq. \eqref{eff_score_V} of this paper comes directly from Eq. (53) of \cite{For_SCRB}. Even if clearly related, the main difference between these two expressions is in the fact that, while in Eq. (53) of \cite{For_SCRB} the gradient is taken w.r.t. $\vsigmat$ where $\bs{\Sigma}_0$ is the \textit{unconstrained} scatter matrix, in this paper the gradient is taken w.r.t. $\ovecs{\mb{V}_1}$ where $\mb{V}_1$ is the \textit{constrained} shape matrix s.t. $[\mb{V}_1]_{11} = 1$. This is the reason why we have the matrix $\mb{M}_N$ instead of the duplication matrix $\mb{D}_N$ as in Eq. (53) of \cite{For_SCRB}. Moreover, Eq. \eqref{eff_score_V} follows from Eq. (53) of \cite{For_SCRB} through basic matrix algebra and the fact that $\trace{\mb{u}_l\mb{u}_l^\top}=\norm{\mb{u}_l}^2=1,\forall l$ and allows us to write a more compact expression for $\mb{s}_{\ovecs{\mb{V}_1},g_0}$.

The efficient SFIM $\bar{\mb{I}}(\bs{\phi}|g_0)$ in \eqref{S_E_FIM} can be immediately obtained from the results in \eqref{eff_score_mu} and \eqref{eff_score_V} and from the expression given in \cite[Eq. (54)]{For_SCRB} as:
\be
\label{E_SFIM}
\begin{split}
	\bar{\mb{I}}(\bs{\phi}|g_0)&\triangleq E_{\bs{\phi},g_0}\{\bar{\mb{s}}_{\bs{\phi},g_0}(\mb{x})\bar{\mb{s}}_{\bs{\phi},g_0}(\mb{x})^\top\} \\
	&= \left( \begin{array}{cc}
		\bar{\mb{I}}(\bs{\mu}|g_0) & \mb{0} \\
		\mb{0}^T   &  \bar{\mb{I}}({\ovecs{\mb{V}_1}}|g_0).
	\end{array}\right),
\end{split}
\ee 
The block-diagonal structure of $\bar{\mb{I}}(\bs{\phi}|g_0)$ in \eqref{E_SFIM} implies that a lack of \textit{a priori} knowledge about the mean vector $\bs{\mu}$ does not have any impact on the asymptotic performance of an estimator of the shape matrix $\mb{V}_1$. In other words, the estimate of $\bs{\mu}$ and the one of $\mb{V}_1$ are asymptotically decorrelated. This and the above-mentioned fact that $\bar{\mb{s}}_{\bs{\mu},g_0} \perp \mathcal{T}_{g_0}$ allow us to considered the estimation of $\bs{\mu}$ and the one of $\mb{V}_1$ as two separate problems. 
%In particular, we do not need an ad-hoc semiparametric procedure to estimate the mean vector $\bs{\mu}$ and we may use any $\sqrt{L}$-consistent estimators. 
For this reason, from now on, we will focus our attention only on the estimation of $\mb{V}_1$.

From \eqref{eff_score_V} and building upon the expression already derived in Eq. (56) of \cite{For_SCRB}, we have that:
\be
\label{cov_mat_eff_scat}
\bar{\mb{I}}({\ovecs{\mb{V}_1}}|g_0) = \alpha_0 \mb{K}_{\mb{V}_1} \mb{K}_{\mb{V}_1}^\top,\; \mathrm{where}
\ee
\be\label{alpha_t}
\alpha_0 \triangleq \nicefrac{2E\{\mathcal{Q}^2\psi_0(\mathcal{Q})^2\}}{N(N+2)}
%\alpha_0 \triangleq 2E\{\mathcal{Q}^2\psi_0(\mathcal{Q})^2\}/N(N+2)
\ee
By substituting the expression of $\bar{\mb{s}}_{\ovecs{\mb{V}_1},g_0}$ given in \eqref{eff_score_V} in the definition of the efficient central sequence in \eqref{cent_seq_sem}, we get:
\be\label{cent_seq_V}
\overline{\bs{\Delta}}_{\mb{V}_1,g_0} = - L^{-1/2}\mb{K}_{\mb{V}_1}\sum\nolimits_{l=1}^{L}Q_l\psi_0(Q_l)  \mathrm{vec}(\mb{u}_l\mb{u}_l^\top).
\ee
Finally, we just need to put \eqref{cent_seq_V} and the expression of $\bar{\mb{I}}({\ovecs{\mb{V}_1}}|g_0)$, given in \eqref{cov_mat_eff_scat}, in the definition of one-step estimator in \eqref{one_step_par_semi}. This yields the following estimator:
\be
\label{semipar_est}
\begin{split}
	\ovecs{\widehat{\mb{V}}_{1,s}} &= \ovecs{\widehat{\mb{V}}_1^\star} -\frac{1}{L\alpha_0}\quadre{\mb{K}_{\widehat{\mb{V}}_1^\star} \mb{K}_{\widehat{\mb{V}}_1^\star}^\top}^{-1} \times \\
	  \mb{K}_{\widehat{\mb{V}}_1^\star}&\sum\nolimits_{l=1}^{L}\hat{Q}^\star_l\psi_0(\hat{Q}^\star_l) \mathrm{vec}(\hat{\mb{u}}^\star_l(\hat{\mb{u}}^\star_l)^\top),
\end{split}
\ee
where:
\be\label{Q_star}
\hat{Q}^\star_l \triangleq (\mb{x}_l-\widehat{\bs{\mu}}^\star)^\top[\widehat{\mb{V}}^\star_1]^{-1}(\mb{x}_l-\widehat{\bs{\mu}}^\star),
\ee
\be\label{u_star}
\hat{\mb{u}}^\star_l \triangleq (\hat{Q}^\star_l)^{-1/2}[\widehat{\mb{V}}^\star_1]^{-1/2}(\mb{x}_l-\widehat{\bs{\mu}}^\star),
\ee
while, as the notation suggests, the matrix $\mb{K}_{\widehat{\mb{V}}_1^\star}$ is obtained from $\mb{K}_{\mb{V}_1}$ in \eqref{K_mat} by substituting $\mb{V}_1$ with its preliminary estimator $\widehat{\mb{V}}_1^\star$.

The last thing to do is to choose preliminary estimators for the mean vector and for the shape matrix. To this end, we can use the joint Tyler's shape and mean vector estimator \cite[Eq. (6)]{joint_robust_M_est}, i.e. $\hat{\bs{\mu}}^\star = \hat{\bs{\mu}}_{Ty}$ and $\widehat{\mb{V}}_1^\star=\widehat{\mb{V}}_{1,Ty}$ with the constraint $[\widehat{\mb{V}}_{1,Ty}]_{11}=1$. This is a good choice since such $\hat{\bs{\phi}}^\star$ is $\sqrt{L}$-consistent under any possible density generator $g \in \mathcal{G}$.

As previously said, the clairvoyant estimators provided in Eq. \eqref{semipar_est} cannot be directly exploited for semiparametric inference since it still depends on the true density generator $g_0$ from two different standpoints:
\begin{itemize}
	\item[\textit{i})] \textit{Statistical dependence}: The estimator $\widehat{\mb{V}}_{1,s}$ in \eqref{semipar_est} relies on the random variables $\{\hat{Q}^\star_l\}_{l=1}^L$ whose pdf depends on $g_0$ through the one of the data $\{\mb{x}_l\}_{l=1}^L$ (see Eq. \eqref{Q_star}).  
	\item[\textit{ii})] \textit{Functional dependence}: The scalar $\alpha_0$ in \eqref{alpha_t} is function of $E\{\mathcal{Q}^2\psi_0(\mathcal{Q})^2\}$ that depends on $g_0$ through the function $\psi_0$ in \eqref{psi} and the pdf of $\mathcal{Q}$ in \eqref{SS_Q_pdf}.
\end{itemize}

In \cite{Hallin_Annals_Stat_2}, Hallin, Oja and Paindaveine showed that rank-based statistics can be exploited to overcome the above-mentioned dependences and obtain a distributionally robust estimator of the shape matrix able to dispense with the knowledge of $g_0$. However, to fully understand the theory underlying the outcomes of \cite{Hallin_Annals_Stat_2}, a strong knowledge of the Le Cam theory and of its \textit{invariance-based} extension to semiparametric framework \cite{Hallin_Werker} is required. The aim of the following subsections is then to supply any SP practitioner with a \virg{ready-to-use} formulation of the resulting $R$-estimator. Anyway, the interested reader can find additional tutorial-style discussions about the semiparametric extension of the Le Cam's theory in the supporting material of this paper.

\subsection{Preliminaries on rank-based statistics} 

%The \textit{ranks} of a set of relevant random variables are a useful tool in non-parametric statistics and numerous works can be found on this topic (see \cite{hajek1968}, \cite[Ch. 13]{vaart_1998} and references therein). 
%Far be it from us to propose a comprehensive overview of the use of ranks in robust statistics, in the following we limit ourselves to introduce their definition and some of their main properties.
Let $\{x_l\}_{l=1}^L$ be a set of $L$ continuous i.i.d. random variables s.t. $x_l \sim p_X,\forall l$. We define the vector of the \textit{order statistics} as $\mb{v}_X \triangleq [x_{L(1)}, x_{L(2)},\ldots,x_{L(L)}]^\top$ whose entries $x_{L(1)}<x_{L(2)}< \cdots < x_{L(L)}$ are the values of $\{x_l\}_{l=1}^L$ ordered in an ascending way.\footnote{Note that, since $x_l,\forall l$ are continuous random variable the equality occurs with probability 0.} Then, the \text{rank} $r_l \in \mathbb{N}/\{0\}$ of $x_l$ is the position index of $x_l$ in $\mb{v}_X$. Finally, we define $\mb{r}_X \triangleq [r_1,\ldots,r_L]^\top \in \mathbb{N}^L$ as the vector collecting the ranks.
\begin{lemma}\label{lemma_ranks}
	Let $\mathcal{K}$ be the family of score functions \footnote{Even if this can create some ambiguity, we decide to indicate the elements in $\mathcal{K}$ as \virg{score functions} in order to maintain the consistency with the terminology used in classical references about ranks.} $K:(0,1)\rightarrow \mathbb{R}^+$ that are continuous, square integrable and that can be expressed as the difference of two monotone increasing functions. Then, we have:
	\begin{enumerate}
		\item The vectors $\mb{v}_X$ and $\mb{r}_X$ are independent,
		\item Regardless the actual pdf $p_X$, the rank vector $\mb{r}_X$ is uniformly distributed on the set of all $L!$ permutations on $\{1,2,\ldots,L\}$ and $!$ stands for the factorial notation,
		\item For each $l=1,\ldots,L$, we have that $K\tonde{\frac{r_l}{L+1}} = K\tonde{u_l} + o_P(1)$
%		\be
%		K\tonde{\frac{r_l}{L+1}} = K\tonde{u_l} + o_P(1),
%		\ee
		where $K \in \mathcal{K}$ and $u_l \sim \mathcal{U}[0,1]$ is a random variable uniformly distributed in $(0,1)$.  
		%		\item Under some regularity conditions on the kernel function $K_0$, we have that:
		%		\be\label{K_rank}
		%		K_0\tonde{\frac{r_l}{L+1}} = Q_l\psi_0(Q_l) + o_P(1),
		%		\ee  
	\end{enumerate}
\end{lemma} 
\textit{Remark}: The proof can be found in \cite{hajek1968}, \cite[Ch. 13]{vaart_1998}.
%or in \cite[Ch. 3]{Hajek}.     

To understand why Lemma \ref{lemma_ranks} is useful to derive a distributionally robust and semiparametric efficient estimator of the shape matrix we should take a step back. 

\subsection{Robust approximations of $\overline{\bs{\Delta}}_{\mb{V}_1,g_0}$ and $\bar{\mb{I}}({\ovecs{\mb{V}_1}}|g_0)$}\label{sec_app_eff_CS} 
From the stochastic representation in \eqref{SRT_dec}, there is a one-to-one correspondence between a RES distributed observation vector $\mb{x}_l \sim RES_{N}(\bs{\mu},\bs{\Sigma},g_0)$ and the couple $(Q_l,\mb{u}_l)$, where $Q_l\overset{d}{=} \mathcal{Q}$ is defined in \eqref{Q_RES} and whose pdf $p_\mathcal{Q}$ is given in \eqref{SS_Q_pdf}, while $\mb{u} \sim \mathcal{U}(\mathbb{R}S^{N-1})$. Then, Point 2) in the Lemma \ref{lemma_ranks} tells us that the distribution of $\mb{r}_Q$ is invariant w.r.t. the pdf $p_\mathcal{Q}$ in \eqref{SS_Q_pdf} that depends on the actual, and generally unknown, density generator $g_0 \in \mathcal{G}$. This feature is very attractive for robust inference since it allows us to derive rank-based (or $R$-) estimators and tests that are distributionally robust. 
%In particular, the distributionally robust version of the efficient central sequence that we are looking for should rely on $\{(r_l,\mb{u}_l)\}_{l=1}^L$ (instead of $\{(Q_l,\mb{u}_l)\}_{l=1}^L$) as the one in \eqref{cent_seq_V} does. 
Point 3) of Lemma \ref{lemma_ranks} provides us with the missing piece to obtain a distributionally robust approximation of the efficient central sequence $\overline{\bs{\Delta}}_{\mb{V}_1,g_0}$. Specifically, let
\be\label{cdf_Q}
P_{\mathcal{Q},0}(q) \triangleq (\pi/2)^{N/2}\Gamma(N/2)^{-1} \int_{0}^{q} t^{N/2-1} g_0 (t)dt
\ee
be the true, and generally unknown, cdf of 2nd-order modular variates whose pdf is given in \eqref{SS_Q_pdf}. Let us now recall the basic fact that (see e.g. \cite[Th. 2.1.10]{CaseBerg})
\be\label{PQ_u}
P_{\mathcal{Q},0}^{-1}(u_l)=Q_l,\quad u_l \sim \mathcal{U}[0,1],\quad Q_l \sim P_{\mathcal{Q},0}\; \forall l
\ee
where $P_{\mathcal{Q},0}^{-1}$ indicates the inverse function of the cdf. Finally, we have to introduce the \virg{true} score function
\be\label{score_true}
K_0(u) \triangleq -P_{\mathcal{Q},0}^{-1}(u)\psi_0(P_{\mathcal{Q},0}^{-1}(u)), \quad u \in (0,1), 
\ee  
that can be shown to belong to the set $\mathcal{K}$ \cite{PAINDAVEINE_CS}. Note that $K_0$ depends on the true density generator $g_0$ through $\psi_0$ in \eqref{psi} and $P_{\mathcal{Q},0}$ in \eqref{cdf_Q}. From Point 3) of Lemma \ref{lemma_ranks} and by using the relation \eqref{PQ_u} we have
\be\label{K_rank}
K_0\tonde{\frac{r_l}{L+1}} = - Q_l\psi_0(Q_l) + o_P(1).
\ee  
Consequently, substituting \eqref{K_rank} in \eqref{eff_score_V} yields to the following approximation of the efficient central sequence in \eqref{cent_seq_V}:
\be\label{cs_ru}
\overline{\bs{\Delta}}_{\mb{V}_1,g_0} = \frac{1}{\sqrt{L}}\mb{K}_{\mb{V}_1}\sum_{l=1}^{L}K_0\tonde{\frac{r_l}{L+1}}  \mathrm{vec}(\mb{u}_l\mb{u}_l^\top) + o_P(1).
\ee
The expression in \eqref{cs_ru} depends \virg{statistically} only on the ranks $r_l$ and on the random vectors $\mb{u}_l$ whose distributions are invariant w.r.t. the actual RES distribution of the data. However, we still have a functional dependence from $g_0$ due to the score function $K_0$. To get rid of this dependence,  
%we can adopt a common procedure in robust statistics: since $K_0$ is unknown, let us choose another function $K \in \mathcal{K}$ that \virg{approximates} $K_0$ and providing, at the same time, a certain amount of robustness. A classical example is the set of the \textit{power} score functions defined as:
%\be
%\label{set_power}
%\mathcal{K}_a \triangleq \graffe{K_a:(0,1)\rightarrow\mathbb{R}^+|K_a(u)=N(a+1)u^a,a \geq 0}.
%\ee
%The celebrated \textit{sign}, \textit{Wilcoxon}, and \textit{Spearman} score functions belong to $\mathcal{K}_a$ in \eqref{set_power} and are obtained for $a=0$, $a=1$ and $a=2$ \cite{Hallin_PCA}. 
we may adopt a \virg{misspecified approach} \cite{SPM}: \textit{since we do not know which is the true density generator $g_0$, let us build the score function $K_g$ by substituting in \eqref{score_true} a, possibly misspecified, $g \in \mathcal{G}$ instead of the unknown $g_0$.} Consequently, by substituting $\mb{V}_1$ with a consistent preliminary estimator $\widehat{\mb{V}}_1^\star$, a distributionally robust approximation of the efficient central sequence $\overline{\bs{\Delta}}_{\mb{V}_1}$ in \eqref{cent_seq_V} can be obtained as:
\be\label{app_eff_cs}
\widetilde{\bs{\Delta}}_{\widehat{\mb{V}}_1^\star} \triangleq \frac{1}{\sqrt{L}}\mb{K}_{\widehat{\mb{V}}_1^\star}\sum_{l=1}^{L}K_g\tonde{\frac{r_l^\star}{L+1}}  \mathrm{vec}(\hat{\mb{u}}^\star_l(\hat{\mb{u}}^\star_l)^\top),
\ee  
where $r_l^\star$ is the rank of $\hat{Q}^\star_l$ already defined in \eqref{Q_star} and $\hat{\mb{u}}^\star_l$ is given in \eqref{u_star}.
As a useful example of score function $K_g$, we may cite the \textit{van der Waerden} score function $K_{vdW}$. Specifically, $K_{vdW}$ is obtained by assuming a, possibly misspecified, Gaussian distribution for the acquired data. Since, under Gaussianity, the density generator is $g_G(t)=\exp(-t/2)$ and $\mathcal{Q}$ in \eqref{Q_RES} is distributed as a $\chi$-squared random variable with $N$ degrees of freedom, i.e. $\mathcal{Q} \sim \chi^2(N)$, from \eqref{score_true} we have:
\be\label{K_G}
K_{vdW}(u) = \Psi^{-1}(u)/2, \quad u \in (0,1),
\ee
where $\Psi(u)$ indicates the cdf of $\chi^2(N)$. \textcolor{blue}{On the same line, if we assume a $t$-distribution for the collected data, we obtain the score function:
\be\label{K_t}
K_{t_\nu}(u) = \frac{N(N +\nu)F^{-1}_{N,\nu}(u)}{2(\nu + NF^{-1}_{N,\nu}(u))},\quad u \in (0,1),
\ee
where $F_{N,\nu}(u)$ stands for the cdf of a Fisher random variable with $N$ and $\nu \in (0,\infty)$ degrees of freedom, i.e. $F_{N,\nu}$. In particular, the expression of $K_{t_\nu}$ comes form the fact that, under and assumed $t$-distribution, the density generator is $g_{t_\nu}(t) = (1+t/\nu)^{-(\nu+N)/2}$ while $\mathcal{Q}/N\sim F_{N,\nu}$ \cite[Ex. 2.5]{RES_Fang}. Note that, from the properties of the $F$-distribution \cite[Ch. 27]{F_dist}, it follows that $\lim_{\nu \rightarrow \infty} K_{t_\nu}(u) = K_{vdW}(u)$. This is not surprising since it is well known that the $t$-distribution collapses into the Gaussian one as $\nu \rightarrow \infty$. We note, that other possible score function may be built upon the loss functions discussed in \cite{Muma_arx}.}

\textcolor{blue}{As expected, a misspecification of the density generator will bring to a loss in semiparametric efficiency. Remarkably, as we will see in Sec. \ref{numerical}, such performance loss are small, especially if the Gaussian \textit{van der Waerden} score is adopted. A theoretical justification of this surprisingly small loss of efficiency may be related to the so-called \virg{Chernoff-Savage result} for non-parametric $R$-tests \cite{chernoff1958}. Some preliminary investigation towards this direction have been provided in \cite{PAINDAVEINE_CS}, but a comprehensive and in-depth analysis of this phenomenon is still missing. Even if of crucial importance, this aspect falls outside the aims of this paper and it is left to future works.} 

Let us now focus on the efficient SFIM in Eq. \eqref{cov_mat_eff_scat}. In \cite{Hallin_Annals_Stat_2}, it is proved that $\bar{\mb{I}}({\ovecs{\mb{V}_1}}|g_0)$ can be approximated as:
\be\label{app_FIM}
\bar{\mb{I}}({\ovecs{\mb{V}_1}}|g_0) = \hat{\alpha}\mb{K}_{\widehat{\mb{V}}_1^\star} \mb{K}_{\widehat{\mb{V}}_1^\star}^\top + o_P(1),
\ee
where $\hat{\alpha}$ is a \textit{consistent} estimator of $\alpha_0$ in \eqref{alpha_t}. In particular, in \cite[Sec. 4]{Hallin_Annals_Stat_2} it is shown that a possible candidate for $\hat{\alpha}$ is:
\be\label{alpha_hat}
\hat{\alpha} = \nicefrac{\norm{\widetilde{\bs{\Delta}}_{\widehat{\mb{V}}_1^\star + L^{-1/2}\mb{H}^0} - \widetilde{\bs{\Delta}}_{\widehat{\mb{V}}_1^\star}}}{ \norm{ \mb{K}_{\widehat{\mb{V}}_1^\star} \mb{K}_{\widehat{\mb{V}}_1^\star}^\top\ovecs{\mb{H}^0}} },
\ee
where $\mb{H}^0$ may be any symmetric matrix whose first top-left entry is equal to 0, i.e. $[\mb{H}^0]_{1,1}=0$. Therefore, the consistent estimator $\hat{\alpha}$ depends on this \virg{small perturbation} matrix $\mb{H}^0$ that can be considered as an hyper-parameter to be defined by the user. Some consideration on the choice of $\mb{H}^0$ will be provided in Sec. \ref{alg_con} where a numerical analysis of the performance of the proposed shape matrix estimator is presented. Note that the estimator $\hat{\alpha}$ in \eqref{alpha_hat} is only an example of a possible estimator for $\alpha_0$, but other procedures may be adopted as well. In \cite[Sec. 4.2]{Hallin_Annals_Stat_2} for example, an ML-based approach is implemented to derive a consistent and efficient estimator for $\alpha_{0}$. However, such ML-based estimator requires the solution of an optimization problem that may become computationally heavy as the matrix dimension increases.

\textcolor{blue}{We conclude this subsection with an important remark on the distributional robustness of $\widetilde{\bs{\Delta}}_{\widehat{\mb{V}}_1^\star}$ in Eq. \eqref{app_eff_cs} and of the approximation of the SFIM given in Eq. \eqref{app_FIM}. These two terms, needed to build a robust version of the $R$-estimator in \eqref{semipar_est}, depend on four random quantities: the preliminary estimator $\widehat{\mb{V}}_1^\star$, the ranks $r_l^\star$, the vectors $\hat{\mb{u}}^\star_l$ and $\hat{\alpha}$. If, as consistent preliminary estimator, we use a distribution-free estimator as the Tyler's one, it can be easily shown that $r_l^\star$ and $\hat{\mb{u}}^\star_l$ are distribution-free as well. This implies that the \virg{approximated} central sequence $\widetilde{\bs{\Delta}}_{\widehat{\mb{V}}_1^\star}$ is itself distribution-free \cite[Prop. 2.1]{Hallin_Annals_Stat_2}. This is not the case for the estimator $\hat{\alpha}$ in \eqref{alpha_hat}. In fact, even if $\widetilde{\bs{\Delta}}_{\widehat{\mb{V}}_1^\star}$ is distribution-free, this is not true for its \virg{perturbed} version $\widetilde{\bs{\Delta}}_{\widehat{\mb{V}}_1^\star + L^{-1/2}\mb{H}^0}$ as proved in \cite[Prop. 2.1, Point (iv)]{Hallin_Annals_Stat_2}. Consequently, the resulting $R$-estimator will not be fully distribution-free. However, it still remain distributionally robust, since $\hat{\alpha}$ is proven to be a consistent estimator of $\alpha_0$ for every possible density generator $g \in \mathcal{G}$ \cite[Sec. 4]{Hallin_Annals_Stat_2}.}     
   
\subsection{The final expression for the real-valued $R$-estimator}\label{subsec_D}
The desired $R$-estimator of real-valued shape matrices in RES distributed data can then be obtained from the the expression of the semiparametric one-step estimator in Theorem \ref{theo_one_step_semipar} by replacing the efficient central sequence $\overline{\bs{\Delta}}_{\hat{\bs{\phi}}^\star,g_0}$ and the efficient SFIM $\bar{\mb{I}}({\ovecs{\mb{V}_1}}|g_0)$ with their approximations provided in Eqs. \eqref{app_eff_cs} and \eqref{app_FIM}, respectively. In particular, a distributionally robust, one-step estimator of $\mb{V}_1$ is given by: 
\be
\label{semipar_R_est}
\begin{split}
	\ovecs{\widehat{\mb{V}}_{1,R}} = \ovecs{\widehat{\mb{V}}_1^\star} &+\frac{1}{L\hat{\alpha}}\quadre{\mb{K}_{\widehat{\mb{V}}_1^\star} \mb{K}_{\widehat{\mb{V}}_1^\star}^\top}^{-1} \times\\
	 \mb{K}_{\widehat{\mb{V}}_1^\star}\sum\nolimits_{l=1}^{L}&K_g\tonde{\frac{r_l^\star}{L+1}} \mathrm{vec}(\hat{\mb{u}}^\star_l(\hat{\mb{u}}^\star_l)^\top),
\end{split}
\ee
where $\{r_l^\star\}_{l=1}^L$ are the ranks of the random variables $\{\hat{Q}_l^\star\}_{l=1}^L$ defined in Eq. \eqref{Q_star}, while $\hat{\mb{u}}^\star_l$ is defined in Eq. \eqref{u_star}. Again, as preliminary estimator of the (constrained) shape matrix we may use the Tyler's estimator $\widehat{\mb{V}}_1^\star=\widehat{\mb{V}}_{1,Ty}$.

\textcolor{blue}{Before moving on, one last comment is in order. It is immediate to verify from the expressions of $\widehat{\mb{V}}_{1,R}$ and $\hat{\alpha}$, given in Eqs. \eqref{semipar_R_est} and \eqref{alpha_hat} respectively, that the $R$-estimator, as function of the score $K_g$, satisfies the following homogeneity property:  $\widehat{\mb{V}}_{1,R}(cK_g)=\widehat{\mb{V}}_{1,R}(K_g)$ for every positive scalar $c>0$. However, if a different estimator of $\alpha_0$ is adopted, this may not be the case and the score should be normalized, e.g. as $\int_{0}^{1}K_g(u)=N$ \cite[Assumption S3]{Hallin_PCA}.}

\section{Extension to Complex ES distributions}
\label{complex_ext}
Building upon the previously obtained results, this section aims at providing an extension of the $R$-estimator in \eqref{semipar_R_est} to the complex-valued shape matrix estimation problem in CES-distributed data. As already shown in \cite{Esa}, \cite[Ch. 4]{book_zoubir} and \cite[Def. II.1]{For_SCRB_complex}, there exists a one-to-one mapping between the set of the CES distributions and a subset of the RES ones. This implies that the theory already developed for the real-valued case can be applied straight to complex-valued data. However, the use of a real representation of complex quantities usually leads to a loss in the clarity and even in the “interpretability” of the results. \textcolor{blue}{This is because the entries of the complex parameter vector are \virg{scrambled} by the $\mathbb{C}\rightarrow \mathbb{R}^2$ mapping and the analysis of the statistical properties of the resulting real version of the estimator may be quite cumbersome. This problem is even more serious when we have to estimate a complex matrix where, in addition to the \virg{scrambling} of the real and imaginary parts due to the $\mathbb{C}\rightarrow \mathbb{R}^2$ mapping, we must take care of the row-column ordering. Having a mathematical tool that allows us to operate directly in the complex field enables us to represent the entries of the parameter vector/matrix in a compact way gaining a lot in terms of both interpretability and feasibility of the obtained estimator.} Best practice is then to use the Wirtinger calculus \cite{Bos_Grad, Remmert, Erik,Complex_M}. Basically, the Wirtinger calculus generalizes the concept of complex derivative to non-holomorphic, real-valued functions of complex variables. In our recent paper \cite{For_SCRB_complex}, the Wirtinger calculus has been exploited to derive the SCRB for the joint estimation of the complex-valued location vector and scatter matrix of a set of CES distributed data. In particular, the complex-valued counterparts of the efficient score vector and of the SFIM for shape matrices in CES data have been evaluated in \cite{For_SCRB_complex}. As for the real-valued case, these two quantities are the basic ingredients to derive a complex version of the $R$-estimator in \eqref{semipar_R_est}. Note that, due to the strong similarity between the properties of the CES and RES distributed random vectors, in the following we will mostly reuse the same notation introduced in Section \ref{preliminaries} for the corresponding entities.

\subsection{CES distributed data: a recall}
\label{preliminaries_CES}
Let $\{\mb{z}_l\}_{l=1}^L \in \mathbb{C}^N$ be a set of complex i.i.d. observation vectors. Let $\mathcal{G}_\mathbb{C}$ be the following set of functions $\mathcal{G}_\mathbb{C} = \graffe{ h: \mathbb{R}^{+} \rightarrow \mathbb{R}^{+} | \int_{0}^{\infty}t^{N-1}h(t)dt < \infty, \int p_Zd\mb{z} =1 }$ \cite{Esa}. Moreover, we indicate with $\mathcal{M}_N^\mathbb{C}$ the set of the Hermitian, positive definite, $N \times N$ complex matrices.

Any CES-distributed random vector $\mb{z}_l=\mb{x}_{R,l}+j\mb{x}_{I,l} \sim CES(\bs{\mu}, \bs{\Sigma},h)$ satisfies the properties \cite{Esa},\cite[Sec. II]{For_SCRB_complex}:
\begin{itemize}
	\item $\mb{z}_l \in \mathbb{C}^N$ is CES distributed iff $[\mb{x}_{R,l}^\top,\mb{x}_{I,l}^\top]^\top \in \mathbb{R}^{2N}$ has a $2N$-variate RES distribution, 
	\item Its pdf $p_Z$ is fully specified by the location vector $\bs{\mu} \in \mathbb{C}^N$, by the scatter matrix $\bs{\Sigma} \in \mathcal{M}_N^\mathbb{C}$ and by the density generator $h \in \mathcal{G}_\mathbb{C}$ and it can be expressed as: 
	\be
	\label{CES_pdf}
	p_Z(\mb{z}_l|\bs{\mu},\bs{\Sigma},h)=|\bs{\Sigma}|^{-1} h \left((\mb{z}_l-\bs{\mu})^\mathsf{H}\bs{\Sigma}^{-1}(\mb{z}_l-\bs{\mu}) \right).
	\ee
	\item \textit{Stochastic representation}: $\mb{z}_l \overset{d}{=} \bs{\mu} + \mathcal{R}\bs{\Sigma}^{1/2}\mb{u}$,
	where $\mathcal{R}$ is the \textit{modular variate} and $\mb{u} \sim \mathcal{U}(\mathbb{C}S^{N-1})$ is uniformly distributed on \textcolor{blue}{$\mathbb{C}S^{N-1} \triangleq \{\mb{u}\in \mathbb{C}^N|\norm{\mb{u}}=1\}$}.
	\item The \textit{2nd-order modular variate} $\mathcal{Q} \triangleq \mathcal{R}^2$ is s.t.
	\be
	\label{Q_CES}
	\mathcal{Q}\overset{d}{=} (\mb{z}_l-\bs{\mu})^\mathsf{H}\bs{\Sigma}^{-1}(\mb{z}_l-\bs{\mu})\triangleq Q_l,\forall l,
	\ee
	and it admits a pdf $p_\mathcal{Q}$ of the form:
	\be
	\label{pdf_Q}
	p_{\mathcal{Q}}(q) = \pi^{N}\Gamma(N)^{-1} q^{N-1} h (q).
	\ee    	
\end{itemize}

Exactly as for the real-valued case, the complex scatter matrix $\bs{\Sigma}$ is not identifiable and only a scaled version of it can be estimated. Then, the shape matrix $\mb{V} \triangleq \bs{\Sigma}/s(\bs{\Sigma})$ has to be introduced, where $s(\cdot)$ is a scalar functional on $\mathcal{M}_N^\mathbb{C}$ satisfying conditions A1, A2 and A3 given in Sec. II. As for the real case, among all the possible scale functionals, we choose $s(\bs{\Sigma}) = [\bs{\Sigma}]_{1,1}$ for simplicity.

At first, we need to define the unknown complex-valued parameter vector $\bs{\phi}$ to be estimated. As shown in \cite{For_SCRB_complex} and in analogy with the real-valued case, the estimation of the mean vector and of the shape matrix are asymptotically decorrelated. Consequently, we focus only of the shape matrix estimation from the \virg{centered} data set $\{\mb{z}_l-\hat{\bs{\mu}}^\star\}_{l=1}^L$, where $\hat{\bs{\mu}}^\star$ is any $\sqrt{L}$-consistent estimator of $\bs{\mu}\in \mathbb{C}^N$. According to the basics of the Wirtinger calculus, $\bs{\phi}$ has to be constructed stacking in a single vector the unknown parameters and their complex conjugate \cite{Bos_Grad,Complex_M}. Then, according to the detailed discussion provided in \cite[Sec. III.A]{For_SCRB_complex}, we have that $\bs{\phi} = \ovec{\mb{V}_1}$.

As shown in Theorem \ref{theo_one_step_semipar}, the basic building blocks for a semiparametric efficient estimators are the semiparametric efficient score vector $\bar{\mb{s}}_{\bs{\phi},h_0} \equiv \bar{\mb{s}}_{\ovec{\mb{V}_1},h_0}$ and the efficient SFIM $\bar{\mb{I}}(\ovec{\mb{V}_1}|h_0)$. Both $\bar{\mb{s}}_{\ovec{\mb{V}_1},h_0}$ and $\bar{\mb{I}}(\ovec{\mb{V}_1}|h_0)$ have been already introduced in full details in our previous work \cite{For_SCRB_complex} and their expressions are recalled here for clarity. Let us start by defining the following matrices:
\be
\label{mat_T}
\mb{P} \triangleq \quadre{\mb{e}_2|\mb{e}_3|\cdots| \mb{e}_{N^2}},
\ee
where $\mb{e}_i$ is the $i$-th vector of the canonical basis of $\mathbb{R}^{N^2},$
\be
\label{L_mat}
\mb{L}_{\mb{V}_1} \triangleq \mb{P} \tonde{\kronVtinvmT} \Pi^{\perp}_{\cvec{\mb{I}_N}},
\ee
and $\Pi^{\perp}_{\cvec{\mb{I}_N}}$ has already been defined in \eqref{mat_proj_I}. Then, from the calculation in \cite[Sec. III.B]{For_SCRB_complex},\footnote{Not that in \cite[Eq. (25)]{For_SCRB_complex} there is a typo. In fact, a minus \virg{$-$} is missing in front of the right-hand side.} using some matrix algebra, we obtain the following expression for the complex efficient semiparametric score vector 
\be
\label{com_eff_score_V}
\bar{\mb{s}}_{\ovec{\mb{V}_1},h_0} = -Q_l\psi_0(Q_l)\mb{L}_{\mb{V}_1}  \mathrm{vec}(\mb{u}_l\mb{u}_l^\mathsf{H}),
\ee
where $\psi_0(t) \triangleq \ln h_0(t)/dt$, $\mb{u}_l \triangleq (Q_l\mb{V}_1)^{-1/2}(\mb{z}_l-\bs{\mu})$
%\be
%\mb{u}_l \triangleq (Q_l\mb{V}_1)^{-1/2}(\mb{z}_l-\bs{\mu}),
%\ee
and $Q_l$ has been defined in \eqref{Q_CES}. Note that the function $\psi_0$ here is defined by means of the true density generator $h_0$ related to the CES pdf in \eqref{CES_pdf}. Moreover, from \cite[Eq. (29)]{For_SCRB_complex}:
\be
\label{com_cov_mat_eff_scat}
\bar{\mb{I}}({\ovec{\mb{V}_1}}|h_0) = \alpha_{\mathbb{C},0}\mb{L}_{\mb{V}_1} \mb{L}_{\mb{V}_1}^\mathsf{H}, \; \mathrm{where}
\ee
\be\label{alpha_t_C}
\alpha_{\mathbb{C},0} \triangleq  \nicefrac{E\{\mathcal{Q}^2\psi_0(\mathcal{Q})^2\}}{N(N+1)}.
\ee
It is worth to underline that the matrix $\mb{P}$ in \eqref{mat_T} has been introduced in order to take into account the fact that the first top-left entry of $\mb{V}_1$ is equal to 1, i.e. $[\mb{V}_1]_{1,1}=1$, and it does not have to be estimated.

\subsection{An $R$-estimator for shape matrices in CES data}
The derivation of the complex-valued $R$-estimator mimics the one proposed in Section \ref{R_est_sec} for the real case. In particular, an approximation of the complex-valued efficient central sequence can be obtained as:
\be\label{complex_app_eff_cs}
\widetilde{\bs{\Delta}}_{\widehat{\mb{V}}_1^\star}^{\mathbb{C}} \triangleq \frac{1}{\sqrt{L}}\mb{L}_{\widehat{\mb{V}}_1^\star}\sum_{l=1}^{L}K_h\tonde{\frac{r_l^\star}{L+1}}  \mathrm{vec}(\hat{\mb{u}}^\star_l(\hat{\mb{u}}^\star_l)^\mathsf{H}),
\ee 
where $\widehat{\mb{V}}_1^\star$ is any $\sqrt{L}$-consistent estimator of the (complex-valued) shape matrix and $r_l^\star$ is the rank of $\hat{Q}^\star_l$ defined, in analogy with \eqref{Q_star}, as
\be\label{CES_Q_star}
\hat{Q}^\star_l \triangleq (\mb{z}_l-\widehat{\bs{\mu}}^\star)^\mathsf{H}[\widehat{\mb{V}}^\star_1]^{-1}(\mb{z}_l-\widehat{\bs{\mu}}^\star),
\ee
\be
\hat{\mb{u}}^\star_l \triangleq (\hat{Q}^\star_l)^{-1/2}[\widehat{\mb{V}}^\star_1]^{-1/2}(\mb{z}_l-\widehat{\bs{\mu}}^\star).
\ee
Moreover, the score function $K_h(\cdot)$ is the \virg{complex} counterpart of the one defined in \eqref{score_true}. Specifically, $K_h(\cdot)$ can be obtained from the expression \eqref{score_true} by evaluating $P_\mathcal{Q}^{-1}$ and $\psi_0$ by means of an assumed, and possibly misspecified, $h \in \mathcal{G}_\mathbb{C}$ instead of its real counterpart $g \in \mathcal{G}$. For example, the \virg{complex version} of the \textit{van der Waerden} score function in \eqref{K_G} can be obtained from \eqref{score_true} by noticing that the complex circular Gaussian distribution has a density generator given by $h_{\mathbb{C}G}(t)=\exp(-t)$ while $\mathcal{Q} \sim \mathrm{Gamma}(N,1)$ \cite{Esa}. Then, the \virg{complex} \textit{van der Waerden} score function is:  
\be\label{K_CG}
K_{\mathbb{C}vdW}(u) \triangleq \Phi_G^{-1}(u),\quad u \in (0,1),
\ee
where $\Phi_G$ indicates the cdf of a Gamma-distributed random variable with parameters $(N,1)$. \textcolor{blue}{Similarly, the \virg{complex version} of the $t_\nu$-score in Eq. \eqref{K_t} is given by:
\be\label{CK_t}
K_{\mathbb{C}t_\nu}(u) = \frac{N(2N+\nu)F^{-1}_{2N,\nu}(u)}{\nu + 2NF^{-1}_{2N,\nu}(u)},\quad u \in (0,1),
\ee
where, as in \eqref{K_t}, $F_{2N,\nu}(u)$ stands for the Fisher cdf with $2N$ and $\nu \in (0,\infty)$ degrees of freedom, where we used the fact that $h_{\mathbb{C}G}(t)=(1+2t/\nu)^{-(2N+\nu)/2}$ and $\mathcal{Q}/N \sim F_{2N,\nu}$ \cite{Esa}. We note that, as for the real case previously discussed, we have that $\lim_{\nu \rightarrow \infty} K_{\mathbb{C}t_\nu}(u) = K_{\mathbb{C}vdW}(u)$.} The complex-valued approximation of the efficient SFIM in \eqref{com_cov_mat_eff_scat} can be obtained as:
\be\label{app_CFIM}
\bar{\mb{I}}({\ovec{\mb{V}_1}}|h_0) = \hat{\alpha}_{\mathbb{C}}\mb{L}_{\widehat{\mb{V}}_1^\star} \mb{L}_{\widehat{\mb{V}}_1^\star}^\mathsf{H} + o_P(1), \; \mathrm{where}
\ee
\be\label{com_alpha_hat}
\boxed{\hat{\alpha}_\mathbb{C} = \nicefrac{\norm{\widetilde{\bs{\Delta}}^\mathbb{C}_{\widehat{\mb{V}}_1^\star + L^{-1/2}\mb{H}^0_\mathbb{C}} - \widetilde{\bs{\Delta}}^\mathbb{C}_{\widehat{\mb{V}}_1^\star}}}{ \norm{ \mb{L}_{\widehat{\mb{V}}_1^\star} \mb{L}_{\widehat{\mb{V}}_1^\star}^\mathsf{H}\ovec{\mb{H}^0_\mathbb{C}}} },}
\ee
and $\mb{H}^0_\mathbb{C}$ is a \virg{small perturbation}, Hermitian, matrix s. t. $[\mb{H}^0_\mathbb{C}]_{1,1}=0$. Finally, putting together the previous results, the complex extension of the distributionally robust, one-step estimator in Eq. \eqref{semipar_R_est} can be obtained as: 
\be
\label{com_one_step_R}
\boxed{\begin{split}
	\ovec{\widehat{\mb{V}}_{1,R}} & = \ovec{\widehat{\mb{V}}_1^\star} +\frac{1}{L\hat{\alpha}_\mathbb{C}}\quadre{\mb{L}_{\widehat{\mb{V}}_1^\star} \mb{L}_{\widehat{\mb{V}}_1^\star}^\mathsf{H}}^{-1} \times\\
	&\mb{L}_{\widehat{\mb{V}}_1^\star}\sum\nolimits_{l=1}^{L}K_h\tonde{\frac{r_l^\star}{L+1}} \mathrm{vec}(\hat{\mb{u}}^\star_l(\hat{\mb{u}}^\star_l)^\mathsf{H}).
\end{split}}
\ee
%where $\{r_l^\star\}_{l=1}^L$ are the ranks of the random variables $\{\hat{Q}_l^\star\}_{l=1}^L$ defined, in analogy with \eqref{Q_star}, as
%\be\label{CES_Q_star}
%\hat{Q}^\star_l \triangleq (\mb{z}_l-\widehat{\bs{\mu}}^\star)^\mathsf{H}[\widehat{\mb{V}}^\star_1]^{-1}(\mb{z}_l-\widehat{\bs{\mu}}^\star),
%\ee
%\be
%\hat{\mb{u}}^\star_l \triangleq (\hat{Q}^\star_l)^{-1/2}[\widehat{\mb{V}}^\star_1]^{-1/2}(\mb{z}_l-\widehat{\bs{\mu}}^\star).
%\ee
In the following, the pseudocode to implement the proposed $R$-estimator is reported, while its related Matlab and Python code can be found at \cite{Code_R}. A good preliminary estimator of the constrained, complex-valued shape matrix, may be Tyler's estimator $\widehat{\mb{V}}_1^\star=\widehat{\mb{V}}_{1,Ty}$.
\begin{algorithm} 
	\caption{Semiparametric efficient $R$-estimator for $\mb{V}_1$}\label{alg_sec}
	\begin{algorithmic}[1]
		\renewcommand{\algorithmicrequire}{\textbf{Input:}}
		\renewcommand{\algorithmicensure}{\textbf{Output:}}
		\REQUIRE $\mb{z}_1,\ldots,\mb{z}_L$; $\widehat{\bs{\mu}}^\star$; $\widehat{\mb{V}}^\star_1$; $K_h(\cdot)$; $\mb{H}^0_\mathbb{C}$.
		\ENSURE  $\widehat{\mb{V}}_{1,R}$.
		%\\ \textit{Data centering}: $\{\mb{z}_l\}_{l=1}^L \leftarrow \{\mb{z}_l-\hat{\bs{\mu}}^\star\}_{l=1}^L$,
		\FOR {$l = l$ to $L$} 
		\STATE $\hat{Q}^\star_l \leftarrow (\mb{z}_l-\widehat{\bs{\mu}}^\star)^\mathsf{H}[\widehat{\mb{V}}^\star_1]^{-1}(\mb{z}_l-\widehat{\bs{\mu}}^\star)$,
		\STATE $\hat{\mb{u}}^\star_l \leftarrow (\hat{Q}^\star_l)^{-1/2}[\widehat{\mb{V}}^\star_1]^{-1/2}(\mb{z}_l-\widehat{\bs{\mu}}^\star)$,
		\ENDFOR
		\STATE Evaluate the ranks $\{r_1^\star,\ldots,r_L^\star\}$ of $\{\hat{Q}^\star_1,\ldots,\hat{Q}^\star_L\}$,
		\STATE $\mb{L}_{\widehat{\mb{V}}^\star_1} \leftarrow \mb{P} ([\widehat{\mb{V}}^\star_1]^{-T/2}\otimes[\widehat{\mb{V}}^\star_1]^{-1/2}) \Pi^{\perp}_{\cvec{\mb{I}_N}}$, 
		\STATE $\widetilde{\bs{\Delta}}_{\widehat{\mb{V}}^\star_1}^{\mathbb{C}} \leftarrow L^{-1/2}\mb{L}_{\widehat{\mb{V}}^\star_1}\sum_{l=1}^{L}K_h\tonde{\frac{r_l^\star}{L+1}}  \mathrm{vec}(\hat{\mb{u}}^\star_l(\hat{\mb{u}}^\star_l)^\mathsf{H})$,
		\STATE Evaluate $\widetilde{\bs{\Delta}}^\mathbb{C}_{\widehat{\mb{V}}_1^\star + L^{-1/2}\mb{H}^0_\mathbb{C}}$ following step 7 with $\widehat{\mb{V}}^\star_1 \leftarrow \widehat{\mb{V}}_1^\star + L^{-1/2}\mb{H}^0_\mathbb{C}$,
		%\\ \textit{LOOP Process}
		\STATE Evaluate $\hat{\alpha}_\mathbb{C}$ as in \eqref{com_alpha_hat}.
		%\STATE $\widehat{\bs{\Upsilon}}_\mathbb{C} \leftarrow \hat{\alpha}_\mathbb{C}\mb{L}_{\widehat{\mb{V}}_1^\star} \mb{L}_{\widehat{\mb{V}}_1^\star}^\mathsf{H}$,
		\STATE $\ovec{\widehat{\mb{V}}_{1,R}} \leftarrow \ovec{\widehat{\mb{V}}_1^\star} + L^{-1/2}[ \hat{\alpha}_\mathbb{C}\mb{L}_{\widehat{\mb{V}}_1^\star} \mb{L}_{\widehat{\mb{V}}_1^\star}^\mathsf{H}]^{-1}\widetilde{\bs{\Delta}}^\mathbb{C}_{\widehat{\mb{V}}_1^\star}$,
		\STATE Reshape $\ovec{\widehat{\mb{V}}_{1,R}}$ in a $N \times N$ matrix with $[\widehat{\mb{V}}_{1,R}]_{1,1}=1$.
		\RETURN $\widehat{\mb{V}}_{1,R}$ 
	\end{algorithmic} 
\end{algorithm}  

\section{Numerical analysis}\label{numerical}
\textcolor{blue}{In this section, thorough numerical simulations, we investigate three different aspects of the considered $R$-estimator of shape matrices: \textit{i}) its semiparametric efficiency, \textit{ii}) its robusteness to outliers and \textit{iii}) its algorithmic properties. In the following, we limit ourselves to report the results related to the complex-valued $R$-estimator proposed in Sec. \ref{complex_ext}, while the corresponding analysis of the real-valued case is provided in the supporting material.} 

\textcolor{blue}{In order to distinguish different estimators, each of them will be indicated as $\widehat{\mb{V}}_{1,\gamma}^\varphi$ where $\gamma$ and $\varphi$ specify the estimator at hand as will will see below. For the sake of consistency with the SP literature on scatter matrix estimation, in the figures, we re-normalized $\widehat{\mb{V}}_{1,\gamma}^\varphi$ in order to have $\trace{\widehat{\mb{V}}_{1,\gamma}^\varphi} = N$. According to the discussion on Sec. \ref{preliminaries}, we can define the re-scaled estimator as:
\be\label{re_norm}
\widehat{\mb{V}}_\gamma^\varphi \triangleq N \widehat{\mb{V}}_{1,\gamma}^\varphi/\trace{\widehat{\mb{V}}_{1,\gamma}^\varphi}.
\ee}
Plotting the MSE of this re-scaled estimator will allow us to underline the fact that the semiparametric efficiency property of the derived $R$-estimator does not depend on the particular scale functional adopted.   
As a reference, in the figures we also report the Constrained Semiparametric CRB (CSCRB) derived in closed form in \cite{For_SCRB_complex}. As performance index for the shape matrix estimators, we use
\be
\varsigma_\gamma^\varphi \triangleq \norm{E\{\mathrm{vec}(	\widehat{\mb{V}}_\gamma^\varphi-\mb{V}_0)\mathrm{vec}(\widehat{\mb{V}}_\gamma^\varphi-\mb{V}_0)^H\}}_F,
\ee
Similarly, as performance bound, we adopt the index:
%\be
%\varepsilon_{CCRB} \triangleq \norm{[\mathrm{CCRB}(\bs{\Sigma}_0)]}_F,
%\ee
\be
\varepsilon_{CSCRB} \triangleq \norm{[\mathrm{CSCRB}(\bs{\Sigma}_0,g_0)]}_F.
\ee
Note that the CSCRB in \cite{For_SCRB_complex} is evaluated for a generic scatter matrix, then we have to chose the constraint accordingly to the definition of the shape matrix at hand (see Sec. \ref{preliminaries}).

We generate the data according to a (true but unknown to the estimators) complex Generalized Gaussian (GG) distribution. The interested reader may find additional simulation related to the complex $t$-distribution in \cite{Eusipco_properties}. The data power is chosen to be $\sigma_X^2 = E_\mathcal{Q}\{\mathcal{Q}\}/N = 4$. Finally, all the numerical indices have been evaluated through $10^6$ Monte Carlo runs. 
The density generator of the complex Generalized Gaussian (GG) distribution is \cite{Esa}:
\be\label{GG_h}
h_0(t) \triangleq \frac{s\Gamma(N)b^{-N/s}}{\pi^{N}\Gamma({N/s)}} \exp \left( -\frac{t^s}{b}\right),\; t\in \mathbb{R}^+ 
\ee
and, according to the value of the shape parameter $s>0$, it can model a distribution with both heavier tails ($0<s<1$) and lighter tails ($s>1$) compared to the Gaussian distribution ($s=1$). \textcolor{blue}{The versatility of the GG distribution is useful to assess the distributional robustness of the proposed $R$-estimator since its properties can be checked in Gaussian, super-Gaussian and sub-Gaussian scenarios.} The setting used in our simulation is as follow:
\begin{itemize}
	\item $\bs{\Sigma}_0$ is a Toeplitz Hermitian matrix whose first column is given by $[1,\rho, \ldots,\rho^{N-1}]^\top$; $\rho = 0.8e^{j2\pi/5}$ and $N=8$.
	\item The \virg{small perturbation} matrix $\mb{H}_\mathbb{C}^0$ is chosen to be a symmetric random matrix s.t. $\mb{H}_\mathbb{C}^0 = (\mb{G}_\mathbb{C}+\mb{G}_\mathbb{C}^\mathsf{H})/2$ where $[\mb{G}_\mathbb{C}]_{i,j} \sim \mathcal{CN}(0,\upsilon^2)$, $[\mb{G}_\mathbb{C}]_{1,1}=0$ and $\upsilon = 0.01$. Note that $\upsilon$ has to be small enough to guarantee that $\widehat{\mb{V}}_1^\star + L^{-1/2}\mb{H}_\mathbb{C}^0 \in \mathcal{M}_N^\mathbb{C}$. \textcolor{blue}{A more exausitve discussion on the choice of $\upsilon$ will be given in Sec. \ref{alg_con}.}
	%\item The \virg{complex} \textit{van der Waerden} score $K_{CG}$ in \eqref{K_CG} is used to build the $R$-estimator.
\end{itemize}

\textcolor{blue}{As previously discussed, the $R$-estimator in Eq. \eqref{com_one_step_R} depends on two \virg{user-defined} quantities: 1) the preliminary estimator $\widehat{\mb{V}}^\star_1$ and 2) the score function $K_h$. In order to assess the impact of their choice on the performance of the $R$-estimator, we perform our simulations by using the Tyler's and the Huber's estimators as preliminary estimators. Moreover, for the Huber's estimator, three different values of the tuning parameter $q$ (i.e. $q=0.9, 0.5, 0.1$) has been adopted \cite[Sec. V.C]{Esa}. Note that the Sample Covariance Matrix (SCM) and Tyler's estimators can be obtained form the Huber's one when $q \rightarrow 1$ and $q \rightarrow 0$, respectively. As score functions, we exploit the \textit{van der Waerden} one given in Eq. \eqref{K_CG} and  the $t_\nu$-score in Eq. \eqref{CK_t} for three different value of $\nu$ ($\nu = 0.1, 1, 5$).}

\subsection{\textcolor{blue}{Semiparametric efficiency}}\label{sec_sem_eff}
\begin{figure}[H]
	\centering
	\includegraphics[height=5cm]{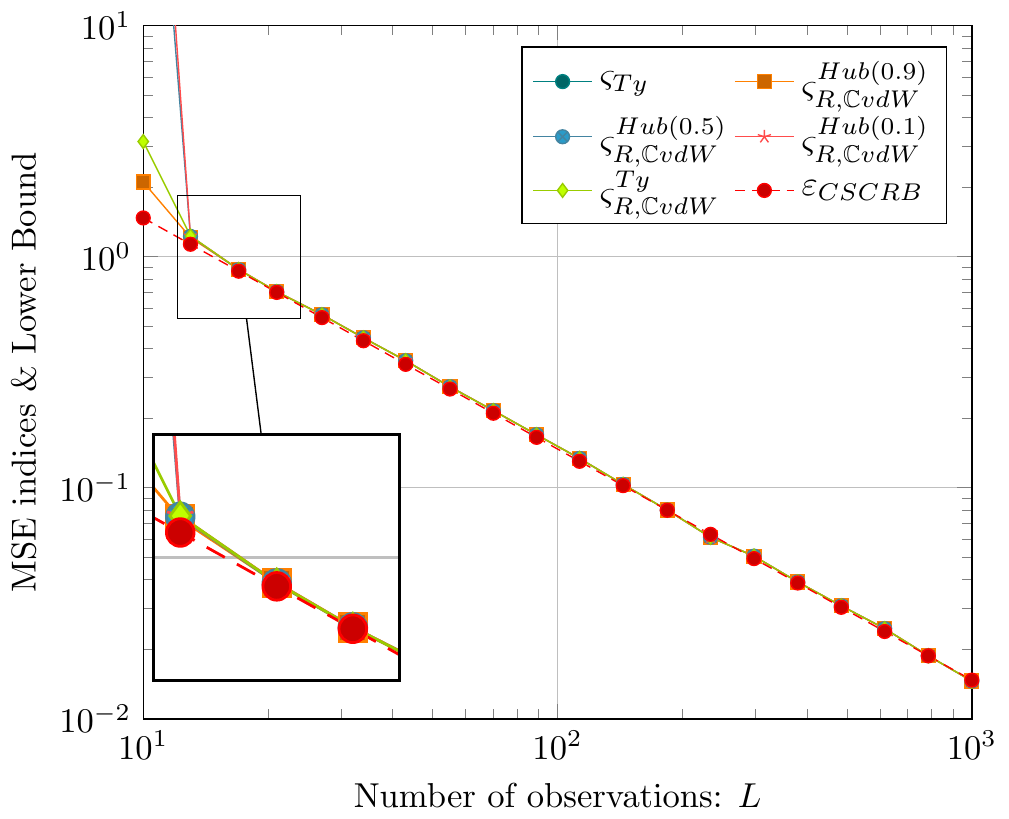}
	\caption{MSE indices vs preliminary Tyler's and Huber's estimators as function of $L$ ($s = 0.5$).}
	\label{fig:Fig_eff_L_pre}
\end{figure}

\begin{figure}[H]
	\centering
	\includegraphics[height=5cm]{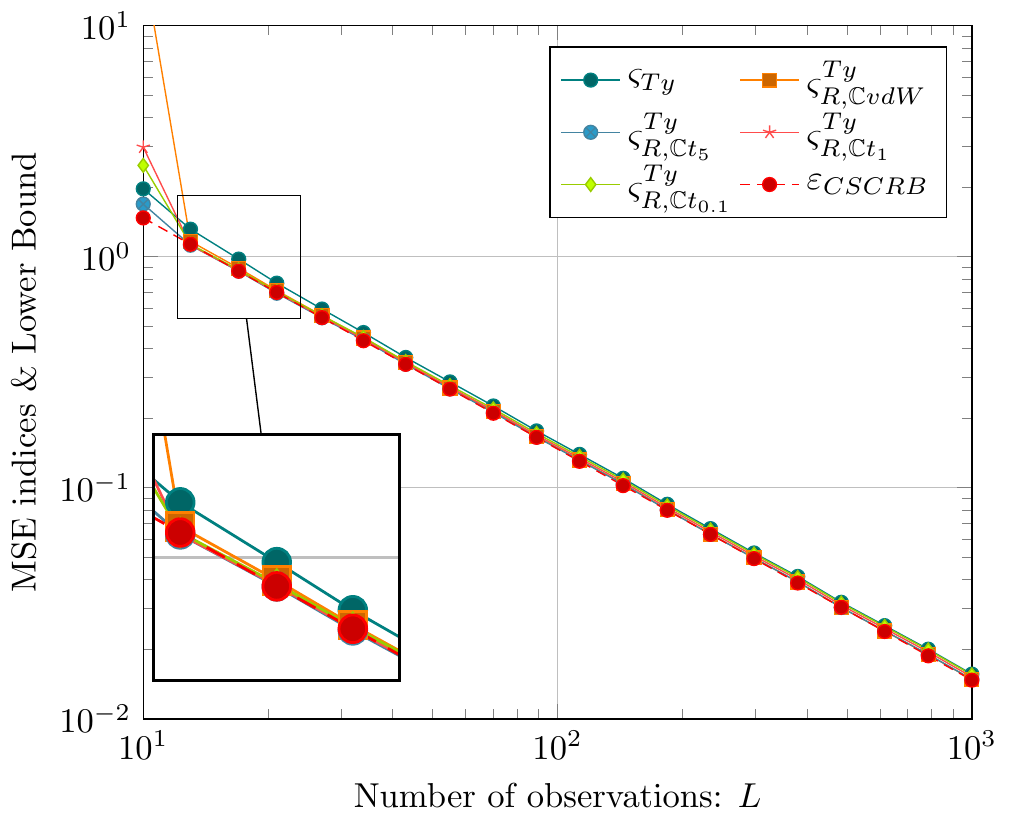}
	\caption{MSE indices vs different score functions $K_h$ as function of $L$ ($s = 0.5$).}
	\label{fig:Fig_eff_L_score}
\end{figure}

\begin{figure}[H]
	\centering
	\includegraphics[height=5cm]{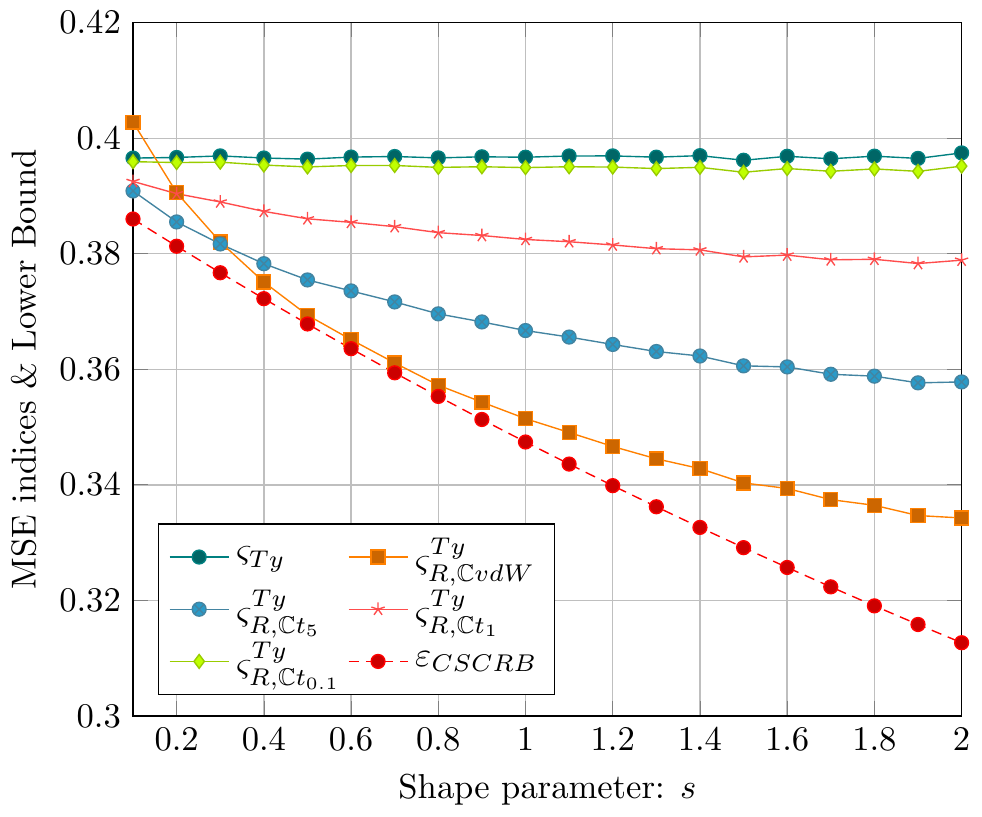}
	\caption{MSE indices vs different score functions $K_h$ as function of $s$ ($L = 5N$).}
	\label{fig:Fig_eff_s_score}
\end{figure}

In Figs. \ref{fig:Fig_eff_L_pre} and \ref{fig:Fig_eff_L_score}, MSE indices of the $R$-estimator in \eqref{com_one_step_R} are plotted as function of the number $L$ of  observations and then compared with the CSCRB for a shape parameter of the GG distribution equal to $0.5$, i.e. for an heavy-tailed scenario. Specifically, in Fig. \ref{fig:Fig_eff_L_pre} the asymptotic efficiency of the $R$-estimator, exploiting a \textit{van der Waerden} score, is investigated for the two considered preliminary estimators, i.e. Tyler's and Huber's one. As we can see, the impact of the choice of the preliminary estimator on the asymptotic efficiency of the $R$-estimator is negligible. Similar consideration can be done for the choice of the particular score function. As shown in Fig. \ref{fig:Fig_eff_L_score} in fact, the MSE curves of the $R$-estimator are very similar to each other and close to the CSCRB as $L\rightarrow \infty$. These simulations confirm the \textit{nearly} semiparametric efficiency of the proposed $R$-estimator. We said \virg{nearly} because, as anticipated in Sec. \ref{sec_app_eff_CS}, the choice of the score function does have an impact on the finite-sample performance and on the robustness to outliers. To see this, in Fig. \ref{fig:Fig_eff_s_score}, we report the MSE indices obtained for the \textit{van der Waerden} and $t_\nu$- scores as function of the shape parameter $s$ in a non-asymptotic regime, i.e. for $L=5N$. The results in Fig. \ref{fig:Fig_eff_s_score} seems to suggest that the \textit{van der Waerden} score provide the lowest MSE index for $0.3<s<2$ while it presents small loss in highly heavy-tailed scenarios ($0.1<s<0.3$). Note that \textit{van der Waerden} score is perfectly specified for $s=1$, i.e. when the data are Gaussian distributed. As anticipated in Sec. \ref{sec_app_eff_CS}, this surprisingly good performance of the \textit{van der Waerden} score is related to the so-called \virg{Chernoff-Savage} result for rank-based statistics \cite{chernoff1958, PAINDAVEINE_CS}. 

The $t_\nu$-scores are more flexible since the additional parameter $\nu$ can be used to tune the desired trade-off between semiparametric efficiency and robustness to outliers, as we will see ahead. In particular, $t_\nu$-scores characterized by a small value of $\nu$ improves the robustness of the resulting $R$-estimator at the price of a loss of efficiency. On the other hand, larger values of $\nu$ will provide a better efficiency, in particular in sub-Gaussian scenario, sacrificing the robustness as addressed in the next section. However, it is important to stress here that the MSE index of the resulting $R$-estimator is lower that the one of Tyler's estimator for all the (non-degenerating) score functions. Moreover, due to the semiparametric nature of the $R$-estimator this conclusion holds true regardless the actual density generator characterizing the data distribution. While the choice of the score function has an impact of the properties of the resulting $R$-estimator, simulation results have highlighted that the impact of the preliminary estimator is negligible, as long as it is $\sqrt{L}$-consistent and robust (see also \cite{Eusipco_properties} for additional discussions). For this reason and for the sake of brevity, in the following we will only report the results obtained by adopting the preliminary Tyler's estimator.              

\subsection{\textcolor{blue}{Robustness to outliers}}
Along with the semiparametric efficiency and distributional robustness, another fundamental property of a shape matrix estimator is the robustness to outliers. In the present context, an outlier is defined as an observation vector that does not share the same statistical behavior of the main data set, i.e. it is not CES distributed or/and it hasn't the same shape matrix or location parameter. The two main tools used to quantify the robustness to outliers of an estimator are the \textit{breakdown point} (BP) and the \textit{influence function} (IF) \cite[Ch. 11 and 12]{huber_book}. Roughly speaking, the BP indicates the percentage of \virg{arbitrary large} outliers that an estimator can tolerate before providing unreliable \virg{arbitrarily large} estimates. On the other hand, the IF gives us a measure of the impact that an infinitesimal perturbation (at a given point) of the samples distribution may have on the estimation performance. Unfortunately, the evaluation of the BP and IF may be involved and difficult to obtain in closed form. Anyway, their \virg{finite-sample} counterparts, called \textit{finite-sample BP} \cite{emp_BP} and \textit{empirical IF} (EIF) \cite{Croux}, or \textit{sensitivity curve}, can be easily evaluated through numerical simulations. 

\begin{figure}[H]
	\centering
	\includegraphics[height=5cm]{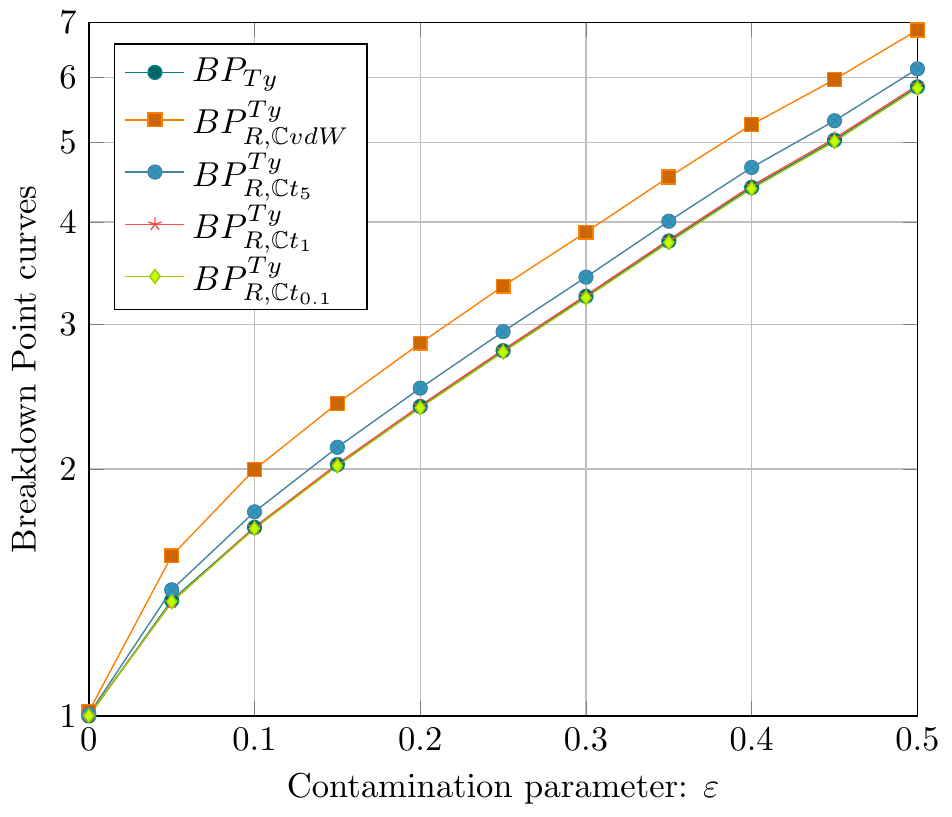}
	\caption{BP vs different score functions $K_h$ as function of $\varepsilon$ ($L = 5N$, $\varrho = 0.1$, $s = 0.1$).}
	\label{fig:Fig_BP_score}
\end{figure}

To evaluate the finite-sample BP for the proposed $R$-estimator, we follow the approach discussed in \cite{BP_Ty}. Let us start by indicating with $Z \triangleq \{\mb{z}_l\}_{l=1}^L \sim CES(\mb{0}, \mb{V}_1,h_0)$ the \virg{pure} GG data set whose $h_0$ is given in \eqref{GG_h} and with $Z_\varepsilon \triangleq \{\mb{z}_{l}\}_{l=1}^L \sim f_{Z_\varepsilon}$ the \textit{$\varepsilon$-contaminated} data set s.t.:
\be\label{cont_model}
f_{Z_\varepsilon}(\mb{z}|\mb{V}_1,h_0,\varrho) = (1-\varepsilon)  CES(\mb{0}, \mb{V}_1,h_0)+ \varepsilon q_Z(\varrho), 
\ee
where $\varepsilon \in [0,1/2]$ is a contamination parameter. The function $q_Z(\varrho)$ represents the pdf of an outlier $\tilde{\mb{z}}$ that we arbitrary choose to be as $\tilde{\mb{z}} = \tau^{-1} \mb{u}$ where, as before, $\mb{u} \sim \mathcal{U}(\mathbb{C}S^{N-1})$ while $\tau \sim \mathrm{Gam}(\varrho,1/\varrho)$ and $\mathrm{Gam}$ indicates the Gamma distribution. Consequently, $\tilde{\mb{z}}|\tau$ is uniformly distributed on the $N$ sphere of ray $\tau^{-1}$, i.e. $\mathbb{C}S^{N-1}_\tau \triangleq \{\tilde{\mb{z}} \in \mathbb{C}^N|\norm{\tilde{\mb{z}}}=\tau^{-1}\}$. This implies that we can obtain \virg{arbitrary large} outlier by generating arbitrary small values of $\tau \sim \mathrm{Gam}(\varrho,1/\varrho)$. This can be achieved by choosing arbitrary small values of the shape parameter $\varrho >0$ in the Gamma distribution. 
%Note that the Gamma distribution has been chosen in analogy to the one of the texture of a $t$-distributed vector \cite{Esa}, but any other positive pdf can be used. 
Let $\widehat{\mb{V}}_\gamma^\varphi(Z)$ and $\widehat{\mb{V}}_\gamma^\varphi(Z_\varepsilon)$ be two shape matrix estimators evaluated from the pure and the $\varepsilon$-contaminated data sets, respectively. Then the finite-sample BP curves can be evaluated as \cite{BP_Ty}:
\be
BP_\gamma^\varphi(\varepsilon) \triangleq \mathrm{max} \graffe{\lambda_{\gamma,1}^\varphi(\varepsilon), 1/\lambda_{\gamma,N}^\varphi(\varepsilon) },
\ee  
where $\lambda_{\gamma,i}^\varphi(\varepsilon)$ is the $i$-th ordered eigenvalue of the matrix $[\widehat{\mb{V}}_\gamma^\varphi(Z)]^{-1}\widehat{\mb{V}}_\gamma^\varphi(Z_\varepsilon)$, s.t. $\lambda_{\gamma,1}^\varphi(\varepsilon) \geq \cdots \geq \lambda_{\gamma,N}^\varphi(\varepsilon)$. Clearly, when there is no contamination ($\varepsilon=0$), we have that $BP_\gamma^\varphi(0) = 1$. Any robust estimator should then have a BP value close to 1 for every value of $\varepsilon$, while it may be arbitrary large for a non-robust estimator. 
%Figs. \ref{fig:Fig_BP_pre} and \ref{fig:Fig_BP_score} show the BP curves of the proposed $R$-estimator exploiting Tyler's and Huber's preliminary estimators (Fig. \ref{fig:Fig_BP_pre}) and for the \textit{van der Waerden} or the $t_\nu$- scores (Fig. \ref{fig:Fig_BP_score}). 
Fig. \ref{fig:Fig_BP_score} shows the BP curves of the proposed $R$-estimator exploiting the \textit{van der Waerden} and three $t_\nu$- scores ($\nu = 0.1, 1, 5$). Since $BP_\gamma^\varphi(\varepsilon)$ depends on $Z$ and $Z_\varepsilon$, we plot its averaged value over $10^4$ realizations of these data sets. For the sake of comparison, we report also the BP value of Tyler's estimator. All the BP curves, related to the resulting $R$-estimator, remain close to the Tyler's one for every value of $\varepsilon$. On the other hand, the BP of the non-robust Sample Covariance Matrix (SCM) estimator explodes to $10^{17}$ as soon as $\varepsilon \neq 0$, so we do not include it in the plot. A visual inspection of Fig. \ref{fig:Fig_BP_score} confirms us what already said in Sec. \ref{sec_sem_eff}: $t_\nu$-scores with a small value of $\nu$ lead to more robust estimators. In particular, it can be noted that the BP curves of the $R$-estimator with $t_{0.1}$- and $t_{1}$-score functions coincide with the one of Tyler's estimator. % is lower of the one of the $t_{1}$- and $t_{5}$- based $R$-estimators. %Note that the \textit{van der Waerden} score provide an intermediate BP curve. %Even if this \virg{finite-sample} analysis seems to suggest that the $R$-estimator have a BP value close to the one of an $M$-estimator of shape \cite{BP_Ty}, more rigorous proof and in-depth investigation of this fact is required and they are left to future works. 

\begin{figure}[H]
	\centering
	\includegraphics[height=5cm]{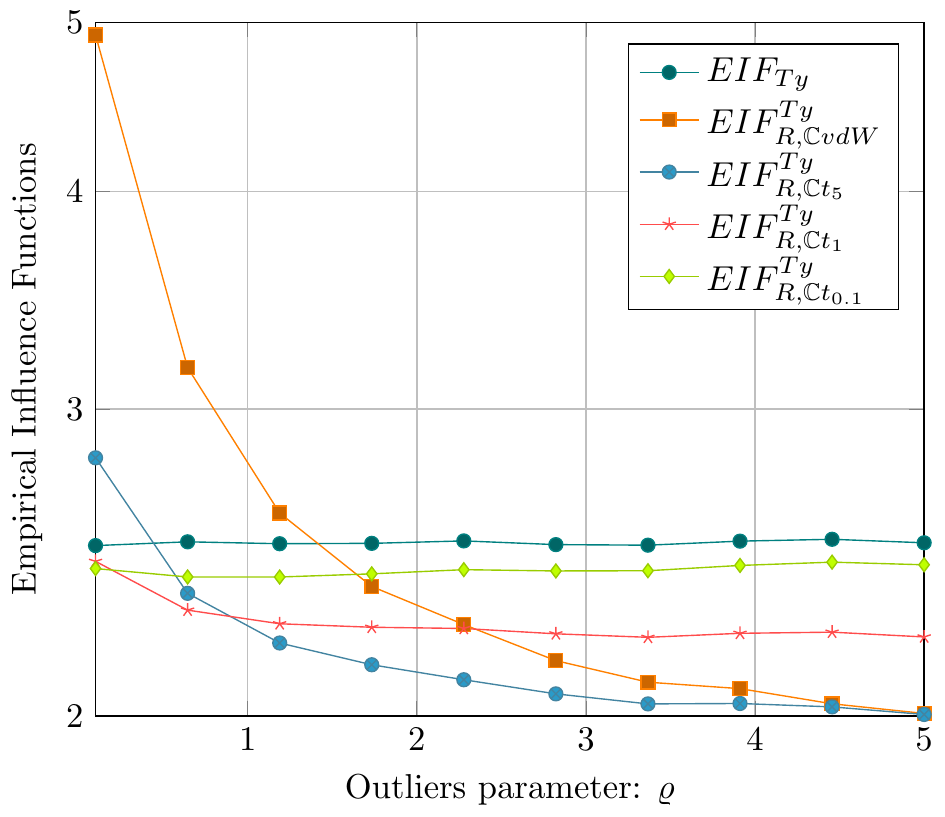}
	\caption{EIF vs different score functions $K_h$ as function of $\varrho$  ($L = 1000$, $s = 0.1$).}
	\label{fig:Fig_SC_score}
\end{figure}

Let us now focus on the EIF \cite{Croux}. For the shape matrix estimation at hand, it can be defined as:
\be\label{sv}
EIF_\gamma^\varphi \triangleq (L+1)\norm{\widehat{\mb{V}}_\gamma^\varphi(Z) - \widehat{\mb{V}}_\gamma^\varphi(Z,\tilde{\mb{z}})}_F,
\ee 
where $\tilde{\mb{z}}$ is an outliers distributed according to the pdf $q_Z(\varrho)$ defined in Eq. \eqref{cont_model}. As Eq. \eqref{sv} suggests, the $EIF_\gamma^\varphi$ gives us a measure of the impact that a \text{single} outlier $\tilde{\mb{z}}$ has on the shape matrix estimator $\widehat{\mb{V}}_\gamma^\varphi$ when it is added to the \virg{pure} data set $Z$. Moreover, if $L$ is sufficiently large, the expression in \eqref{sv} is a good approximation of the theoretical IF \cite{Croux}. For this reason, in our simulation we use $L=1000$. Since $EIF_\gamma^\varphi$ depends on $Z$ and $\tilde{\mb{z}}$, we plot its averaged value over $10^4$ realizations of the data set and the outlier. As for the IF, the most important property that the EIF of a robust estimator should have is the boundeness. In fact, this indicates that the impact of a single outlier on the estimation performance is limited. In Fig. \ref{fig:Fig_SC_score}, we report the EIF of the proposed $R$-estimator exploiting the \textit{van der Waerden} and three $t_\nu$- scores ($\nu = 0.1, 1, 5$). As benchmark, the EIF of the Tyler's estimator is adopted since it is known that the relevant IF is continuous and bounded \cite{Esa}. On the other hand, the EIF of the non-robust SCM grows rapidly to $10^4$ as the norm of the outlier $\tilde{\mb{z}}$ increases (i.e. when $\varrho \rightarrow 0$), so we do not include it in the plot. As we can see from Fig. \ref{fig:Fig_SC_score}, the EIFs of the proposed $R$-estimator remain bounded and close to the one of the Tyler's estimator for arbitrary large vale of $\norm{\tilde{\mb{z}}}$ ($\varrho \rightarrow 0$). 
%Moreover, the results on the robustness induced by the parameter $\nu$ of the $t_\nu$-score are confirmed by the EIF curves: lower values of $\nu$ lead to lower value of the EIF in the presence of an arbitrary large outlier, and consequently, to a more robust $R$-estimator. 
%However, as already said for the BP, this stimulative study provides us with a limited and preliminary information about the rebusteness properties of the $R$-estimator in Eq. \eqref{com_one_step_R}. A more insightful investigation is required and it is left to future works. 

\subsection{\textcolor{blue}{Algorithmic considerations}}\label{alg_con}
This last subsection collects some observations on the algorithmic implementation of the proposed $R$-estimator. As can be seen from the pseudo-code in Sec. \ref{complex_ext}, the $R$-estimator is obtained by applying a linear \virg{one-step} correction $L^{-1/2}[ \hat{\alpha}_\mathbb{C}\mb{L}_{\widehat{\mb{V}}_1^\star} \mb{L}_{\widehat{\mb{V}}_1^\star}^\mathsf{H}]^{-1}\widetilde{\bs{\Delta}}^\mathbb{C}_{\widehat{\mb{V}}_1^\star}$ to a preliminary estimator $\widehat{\mb{V}}_1^\star$ (see step 10 in Algo. \ref{alg_sec}). In particular, unlike $M$-estimators that are obtained as implicit solution of a fixed point equation, it does not require any iterative implementation. Consequently, leaving aside the computation of $\widehat{\mb{V}}_1^\star$, the computational load of the proposed $R$-estimator is roughly given by the amount of calculation needed to \textit{i}) obtain the $L$ ranks $r_l^\star$ and vectors $\hat{\mb{u}}_l^\star$ (see steps 2 and 3 in Algo. \ref{alg_sec}) and \textit{ii}) deal with the $(N^2-1) \times (N^2-1)$ matrices $\mb{L}_{\mb{V}_1}$, $[\mb{L}_{\mb{V}_1} \mb{L}_{\mb{V}_1}^\mathsf{H}]$ and $[\mb{L}_{\mb{V}_1} \mb{L}_{\mb{V}_1}^\mathsf{H}]^{-1}$. Clearly, this represents a problem as the dimension $N$ of the observations increases. A possible way out would be to exploit the structure of $\mb{L}_{\widehat{\mb{V}}_1^\star}$, given in Eq. \eqref{L_mat}, to reduce the global computational load but this point falls outside the scope of the present paper.  

\begin{figure}[H]
	\centering
	\includegraphics[height=5cm]{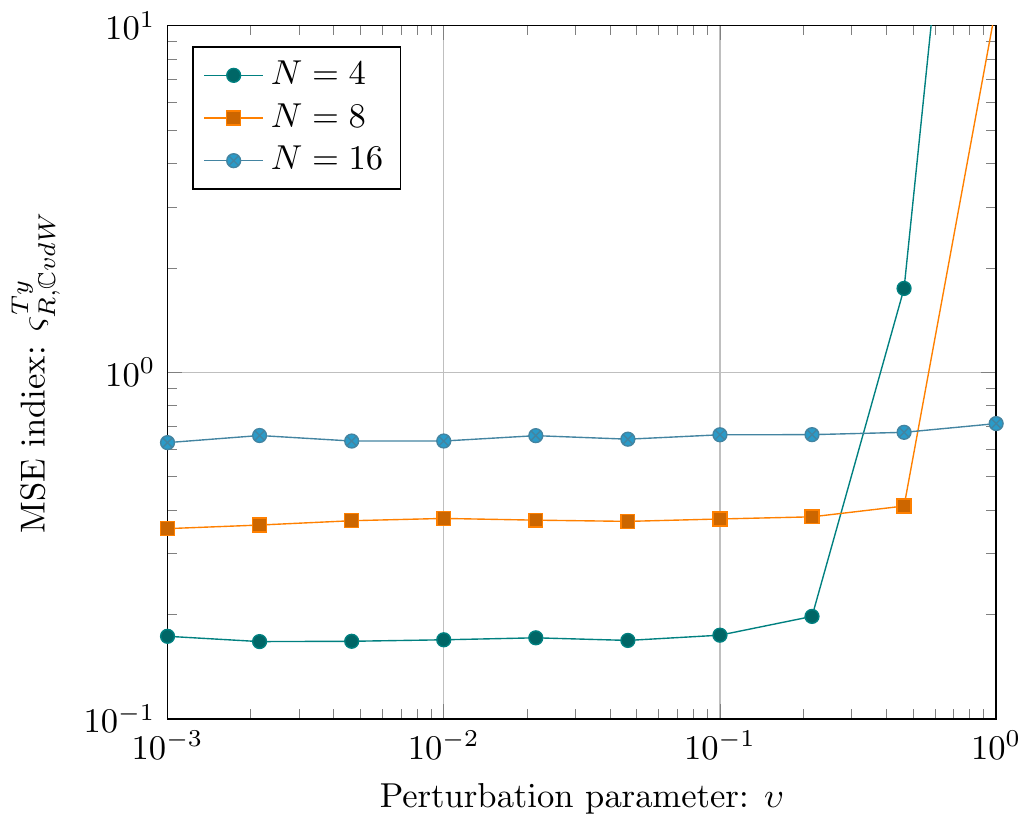}
	\caption{MSE indiex $\varsigma_{R,\mathbb{C}vdW}^{Ty}$ as function of the perturbation parameter $\upsilon$ ($L=5N$, $s = 0.5$).}
	\label{fig:Fig_Per_par}
\end{figure}

The second algorithmic consideration is related the choice of the \virg{small perturbation} matrix $\mb{H}_\mathbb{C}^0$. The theory does not provide us with any hint about the optimal selection of this hyper-parameter, so we decided to define it as a random matrix $\mb{H}_\mathbb{C}^0 = (\mb{G}_\mathbb{C}+\mb{G}_\mathbb{C}^\mathsf{H})/2$ where $[\mb{G}_\mathbb{C}]_{i,j} \sim \mathcal{CN}(0,\upsilon^2)$, $[\mb{G}_\mathbb{C}]_{1,1}=0$. The problem then is reduced to the simpler choice of the scalar perturbation parameter $\upsilon$. Fortunately, simulation results seem to suggest that the $R$-estimator is quite robust w.r.t. the choice of $\upsilon$ for various density generators and various levels of non-Gaussianity. On the other hand, the choice of $\upsilon$ is sensible to the data dimension $N$ and to the number of observations $L$. As an example, Fig. \ref{fig:Fig_Per_par} shows the MSE index of the $\widehat{\mb{V}}_{R,vdW}^{Ty}$ as function of $\upsilon$ for different data dimension $N$. As we can see, the MSE index remains stable for a sufficiently large range of values for $\upsilon$ allowing us for its safe selection.

%The simulation results related to the complex GG-distributed data proposed in Fig. \ref{fig:Fig2} confirm all the observation already done for the real-valued $t$-distributed one. The only difference here is that the $R$-estimator experiences some efficiency loss in the sub-Gaussian case, i.e. for $s>1$. However, it still has better performance w.r.t. the Tyler's estimator for every values of the shape parameter $s$. 
%
%A last comment on the \virg{small perturbation} matrix $\mb{H}^0$ is in order now. The theory, in fact, does not provide us with any hint for the optimal selection of this hyper-parameter. Fortunately, simulation results seem to suggest that the $R$-estimator is quite robust w.r.t. the choice of $\mb{H}^0$ for various density generators and various levels of non-Gaussianity. However, the choice of $\mb{H}^0$ could be sensible to the data dimension $N$ and to the number of observations $L$. We left to future works an in-depth and analysis of this importation aspect. 

\section{Conclusions}
\label{conclusions}
In this paper, a distributionally robust and nearly semiparametric efficient $R$-estimator of the shape matrix in Real and Complex ES distributions has been discussed and analyzed. This estimator has been firstly proposed by Hallin, Oja and Paindaveine in their seminal paper \cite{Hallin_Annals_Stat_2} where the Le Cam's theory of one-step efficient estimators and the properties of rank-based statistics have been exploited as basic building blocks for its derivation. In the first part of this paper, a survey of the main statistical concepts underlying such $R$-estimator has been proposed for the case of RES-distributed data. Then, its extension to CES distributions has been derived by means of the Wirtinger calculus. Finally, the finite-sample performance of the $R$-estimator has been investigated in different scenarios in terms of MSE and robustness to outliers. However, a number of fundamental issues still remain to be fully addressed. In our opinion, the most important one is related to the estimation of $\alpha_{\mathbb{C}, 0}$ in \eqref{alpha_t_C} (or, for the real-valued case,  $\alpha_{0}$ in \eqref{alpha_t}). The estimator in \eqref{com_alpha_hat} in fact is consistent under any possible density generator $h \in \mathcal{G}_\mathbb{C}$ but it does not satisfy any optimality property. Moreover, it depends on an hyper-parameter, i.e. the \virg{small perturbation} matrix $\mb{H}_\mathbb{C}^0$ (or $\mb{H}^0$ in the real-valued case), that has to be defined by the user in an heuristic way and, currently, without any theoretical guidelines. A possible improvement w.r.t. the estimator in \eqref{com_alpha_hat} is discussed in \cite[Sec. 4.2]{Hallin_Annals_Stat_2} and it will be the subject of future works. Other important open questions are related to the evaluation of the theoretical BP point and IF. Closed form expressions of these two quantities will help to fully understand the robustness properties of the proposed $R$-estimator with respect to classical $M$-estimators.

\bibliographystyle{IEEEtran}
\bibliography{ref_semipar_eff_estim}

%\begin{figure}[h]
%	\centering
%	\includegraphics[height=5cm]{images/Eff_s_preliminary.pdf}
%	\caption{MSE indices vs preliminary Tyler's and Huber's estimators as function of $s$ ($L = 5N$).}
%	\label{fig:Fig_eff_s_pre}
%\end{figure}

%\begin{figure}[ht]
%	\centering
%	\includegraphics[height=5cm]{images/BP_preliminary.pdf}
%	\caption{BP vs preliminary Tyler's and Huber's estimators as function of $\varepsilon$ ($L = 5N$, $\varrho = 0.05$, $s = 0.1$).}
%	\label{fig:Fig_BP_pre}
%\end{figure}

%\begin{figure}[h]
%	\centering
%	\includegraphics[height=5cm]{images/SC_preliminary.pdf}
%	\caption{EIF vs preliminary Tyler's and Huber's estimators as function of $\varrho$ ($L = 1000$, $s = 0.1$).}
%	\label{fig:Fig_SC_pre}
%\end{figure}

\onecolumn

	\begin{center}
	
	% MAKE SURE YOU TAKE OUT THE SQUARE BRACKETS
	
	\Large{Supporting material for the paper:} \\
	\vspace{1em}
	\LARGE{Robust Semiparametric Efficient Estimators in Elliptical Distributions} \\
	\vspace{1em}
	\normalsize\textbf{Stefano Fortunati, Alexandre~Renaux, Fr\'{e}d\'{e}ric~Pascal} \\
	
	%\normalsize{E-mail: \href{mailto: stefano.fortunati@iet.unipi.it}{stefano.fortunati@iet.unipi.it}, \href{mailto: fortunati.stefano@gmail.com}{fortunati.stefano@gmail.com} }
	
\end{center}
\begin{normalsize}
	\setcounter{equation}{61}
	
	\section{Le Cam's one-step estimators in a nutshell}
	\label{Introduction1}
	
	The aim of this first section is to provide the reader of our paper with some additional discussion about the general theory of \textit{efficient one-step estimators}. This class of estimators has its root in the concept of  Local Asymptotic Normality (LAN) of a statistical model. The LAN property has been introduced for the first time by Le Cam in his fundamental work \cite{LeCam60} (see also \cite[Ch. 6]{LeCam}) and it has since established itself as a milestone in modern statistics. Leaving aside the deep theoretical implications that the LAN property has for a given family of distributions, there is at least one outcome of great interest for any practitioner working in signal processing (SP) and related fields. As Le Cam showed, if a statistical model is Locally Asymptotic Normal, then it is possible to derive asymptotically efficient estimators that, unlike the Maximum Likelihood (ML) one, do not search for the maxima of the log-likelihood function. This fact is of great importance in practical applications, where the ML estimator can present computational difficulties in the resulting optimization problem or even existence/uniqueness issues \cite[Ch. 6]{Lehm98}. 
	
	We start by introducing the concept of \textit{Hellinger differentiability}, or \textit{differentiability in quadratic mean}. Then, the definition of the LAN property for \textit{parametric models} will be given and its exploitation, in deriving efficient one-step estimators, discussed. Finally, the generalization of the previously developed theory to \textit{semiparametric models} will be provided. 
	
	\textit{Algebraic notation}: Throughout this document, italics indicates scalar quantities ($a$), lower case and upper case boldface indicate column vectors ($\mathbf{a}$) and matrices ($\mathbf{A}$), respectively. Each entry of a matrix $\mb{A}$ is indicated as $a_{ij}\triangleq [\mb{A}]_{i,j}$. $\mb{I}_N$ defines the $N \times N$ identity matrix. The superscript $\top$ indicates the transpose operator, then $\mb{A}^{-\top} \triangleq (\mb{A}^{-1})^\top = (\mb{A}^\top)^{-1}$. The Euclidean norm of a vector $\mb{a}$ is indicated as $\norm{\mb{a}}$. The determinant and the Frobenius norm of a matrix $\mb{A}$ are indicated as $|\mb{A}|$ and $\norm{\mb{A}}_F$, respectively. 
	
	\textit{Small o notation}: Given a real-valued function $f(x)$ and a strictly positive real-valued function $g(x)$, $f(x)=o(g(x))$ if for every positive real number $a$, there exists a real number $x_0$ such that $|f(x)| \leq a g(x),\; \forall x \geq x_0$.

	\textit{Statistical notation}: Let $x_l$ be a sequence of random variables in the same probability space. We write:
	\begin{itemize}
		\item $x_l = o_P(1)$ if $\lim_{l\rightarrow \infty}\mathrm{Pr}\graffe{|x_l|\geq\epsilon}=0,\forall \epsilon>0$ (\textit{convergence in probability to 0}),
		\item $x_l = O_P(1)$ if for any $\epsilon > 0$, there exists a finite $M>0$ and a finite $L>0$, s.t. $\mathrm{Pr}\graffe{|x_l|>M}<\epsilon,\forall l>L$ (\textit{stochastic boundedness}).
	\end{itemize}
	
	The cumulative distribution function (cdf) and the related probability density function (pdf) of a random variable $x$ or a random vector $\mb{x}$ are indicated as $P_X$ and $p_X$, respectively. For random variables and vectors, $\overset{d}{=}$ stands for \virg{has the same distribution as}. The symbol $\underset{L \rightarrow \infty}{\sim}$ indicates the convergence in distribution. We indicate the \textit{true} pdf as $p_0(\mb{x})\triangleq p_X(\mb{x}|\bs{\phi}_0, g_0)$, where $\bs{\phi}_0$ and $g_0$ indicate the true parameter vector to be estimated and the true nuisance function, respectively. We define as $E_{\bs{\phi}, g}\{f(\mb{x})\} = \int f(\mb{x})p_X(\mb{x}|\bs{\phi}, g) d\mb{x}$ the expectation operator of a \textit{measurable} function $f$ of a random vector $\mb{x}$. Moreover, we simply indicate as $E_0\{\cdot\}$ the expectation with respect to (w.r.t.) the true pdf $p_0(\mb{x})$. The superscript $\star$ indicates a $\sqrt{L}$-consistent, \textit{preliminary}, estimator $\hat{\bs{\phi}}^\star$ of $\bs{\phi}_0$, s.t. $\sqrt{L} \tonde{\bs{\phi}^\star - \bs{\phi}_0} = O_{P}(1)$.   
	
	Let $\mb{x} \in \mathbb{R}^N$ be a real-valued random vector and let $p_X$ be its probability density function (pdf). A parametric model, characterizing the statistical behavior of $\mb{x}$, will be indicated as:
	\be
	\label{par_model}
	\mathcal{P}_{\bs{\phi}} = \graffe{p_X | p_X(\mb{x}|\bs{\phi});  \bs{\phi} \in \Omega \subseteq \mathbb{R}^q },
	\ee 
	while a semiparametric model will be described as:
	\be
	\label{semipar_model}
	\mathcal{P}_{\bs{\phi},g} = \graffe{p_X | p_X(\mb{x}|\bs{\phi},g);  \bs{\phi} \in \Omega \subseteq \mathbb{R}^q, g \in \mathcal{G} },
	\ee 
	where $\mathcal{G}$ is a suitable set of functions.
	
	\subsection{Hellinger differentiability}
	Let $\bs{\phi} \in \Omega \subseteq \mathbb{R}^q$ be the parameter vector and let $p_X(\mb{x}|\bs{\phi}) \in \mathcal{P}_{\bs{\phi}}$ be a pdf belonging to the \textit{parametric model} $\mathcal{P}_{\bs{\phi}}$ in \eqref{par_model}. We define $u_{\bs{\phi}}(\mb{x})$ as the following parametric map:
	\be
	\begin{split}
		u_{\bs{\phi}} :\Omega &\to \mathcal{L}_2 \\
		\bs{\phi} &\mapsto u_{\bs{\phi}}(\mb{x})\triangleq \sqrt{p_X(\mb{x}|\bs{\phi})},
	\end{split}
	\ee
	where $\mathcal{L}_2$ indicates the set of all the square integrable functions. We say that $u_{\bs{\phi}}$ is Hellinger differentiable in $\bs{\phi} \in \Omega$ if there exists a vector $\dot{\mb{u}}_{\bs{\phi}} \equiv \dot{\mb{u}}_{\bs{\phi}}(\mb{x})$ such that \cite[Ch. 2, Def. 1]{BKRW}, \cite[Ch. 5.5]{vaart_1998}:
	\be\label{HD}
	\int \quadre{u_{\bs{\phi} + \mb{h}}(\mb{x}) - u_{\bs{\phi}}(\mb{x}) - \mb{h}^\top\dot{\mb{u}}_{\bs{\phi}}(\mb{x})}^2d\mb{x} = o(\norm{\mb{h}}), \quad \mb{h} \in \Omega, \quad \norm{\mb{h}} \rightarrow 0.
	\ee
	Then $\dot{\mb{u}}_{\bs{\phi}} \equiv \dot{\mb{u}}_{\bs{\phi}}(\mb{x})$ is the Hellinger derivative of $u_{\bs{\phi}}$ in $\bs{\phi} \in \Omega$. According to \cite[Ch. 2, Def. 2]{BKRW}, a parametric model $\mathcal{P}_{\bs{\phi}}$ is said to be \textit{regular} if each $p_X(\mb{x}|\bs{\phi}) \in \mathcal{P}_{\bs{\phi}}$ is Hellinger differentiable at every $\bs{\phi} \in \Omega$.
	
	The Hellinger differentiability was introduced by Le Cam as the weakest regularity condition required to develop the LAN theory. However, even if extremely useful for theoretical purposes, the Hellinger differentiability is not really suitable to derive practical inference algorithms. Fortunately, statistical models involved in practical signal processing (SP) applications can generally satisfy more stringent assumptions than the one in \eqref{HD}. This allows us to link the regularity \virg{\textit{\'a la Le Cam}} of a parametric model to more familiar quantities, e.g.\ the \textit{score vector} and the \textit{Fisher Information Matrix} (FIM), as detailed in the following Proposition (see \cite[Ch. 2, Prop. 1]{BKRW} for the proof).
	\begin{proposition}
		\label{Prop_reg}
		Let $\mb{x}$ be a set of $N$-dimensional, real-valued, random vector sampled from a pdf $p_X \in \mathcal{P}_{\bs{\phi}}$ in \eqref{par_model}. Let $\mb{s}_{\bs{\phi}} \equiv \mb{s}_{\bs{\phi}}(\mb{x})$ be the score vector defined as:
		\be
		\label{score_vect1}
		\mb{s}_{\bs{\phi}}(\mb{x})=\nabla_{\bs{\phi}}\ln p_X(\mb{x}|\bs{\phi})
		\ee   
		and let $\mb{I}(\bs{\phi})$ be the Fisher Information Matrix (FIM):
		\be
		\label{FIM}
		\mb{I}(\bs{\phi}) \triangleq E_{\bs{\phi}}\graffe{\mb{s}_{\bs{\phi}}(\mb{x})\mb{s}^\top_{\bs{\phi}}(\mb{x})}.
		\ee
		
		Then, the parametric model $\mathcal{P}_{\bs{\phi}}$ is regular \virg{\textit{\'a la Le Cam}} if the following three sufficient (but not necessary) conditions are satisfied: 
		\begin{itemize}
			\item[\textit{i})] $p_X(\mb{x}|\bs{\phi})$ is continuously differentiable in $\bs{\phi} \in \Omega$ for almost all $\mb{x}$ with gradient $\nabla_{\bs{\phi}} p_X(\mb{x}|\bs{\phi})$,
			\item[\textit{ii})] $E_{\bs{\phi}}\{\mb{s}_{\bs{\phi}}(\mb{x})^\top\mb{s}_{\bs{\phi}}(\mb{x})\} < \infty$,
			\item[\textit{iii})] The FIM in \eqref{FIM} is non-singular and continuous in $\bs{\phi} \in \Omega$.
		\end{itemize}
		If \textit{i}), \textit{ii}) and \textit{iii}) hold true, the Hellinger derivative $\dot{\mb{u}}_{\bs{\phi}}$ defined \eqref{HD} can be explicitly expressed as function of the score vector $\mb{s}_{\bs{\phi}}$ in \eqref{score_vect1} as:
		\be\label{HD_s}
		\dot{\mb{u}}_{\bs{\phi}}(\mb{x}) = \frac{1}{2}\sqrt{p_X(\mb{x}|\bs{\phi})}\mb{s}_{\bs{\phi}}(\mb{x}).
		\ee 
	\end{proposition}
	
	The regularity conditions \textit{i}), \textit{ii}) and \textit{iii}) in Prop. \ref{Prop_reg} requires, among others, the pointwise differentiability of the pdf and consequently they are more stringent than the integral condition in \eqref{HD}. However, they are generally satisfied by the vast majority of the statistical models exploited in practical inference problems. For this reason, in the following discussion, we will assume them for granted but we will always indicate when the obtained results can be derived starting form the weaker regularity condition in \eqref{HD}.
	
	\subsection{LAN property and ES distributions} 
	
	The following Proposition introduces the fundamental LAN property (\cite{LeCam60}, \cite[Ch. 6]{LeCam}, \cite[Ch. 7.6]{vaart_1998}) of a parametric model satisfying the regularity conditions stated in Prop. \ref{Prop_reg}.
	\begin{proposition}
		\label{Prop_LAN}
		Let $\{\mb{x}_l\}_{l=1}^L$ be a set of real-valued, i.i.d. observations sampled from a pdf $p_X$ belonging to a regular parametric model $\mathcal{P}_{\bs{\phi}}$ in \eqref{par_model}. 
		Let $\bs{\Delta}_{\bs{\phi}}(\mb{x}_1,\ldots,\mb{x}_L)$ be a random vector, usually referred to as central sequence, defined as:
		\be\label{cent_seq}
		\bs{\Delta}_{\bs{\phi}}(\mb{x}_1,\ldots,\mb{x}_L) \equiv \bs{\Delta}_{\bs{\phi}} \triangleq L^{-1/2}\sum\nolimits_{l=1}^{L}\mb{s}_{\bs{\phi}}(\mb{x}_l),
		\ee
		where $\mb{s}_{\bs{\phi}}(\mb{x}_l)$ is the score vector given in \eqref{score_vect1}.
		
		Then, any $p_X(\mb{x}|\bs{\phi}) \in \mathcal{P}_{\bs{\phi}}$ satisfies the following LAN property:
		\be
		\label{LAN}
		\ln \frac{\prod\nolimits_{l=1}^L p_X(\mb{x}_l|\bs{\phi}+L^{-1/2}\mb{h})}{\prod\nolimits_{l=1}^L p_X(\mb{x}_l|\bs{\phi})} = \mb{h}^\top\bs{\Delta}_{\bs{\phi}}- \frac{1}{2} \mb{h}^\top\mb{I}(\bs{\phi})\mb{h} + o_P(1), \quad \forall \bs{\phi},\mb{h} \in \Omega,
		\ee 
		where $\mb{I}(\bs{\phi})$ is the FIM given in \eqref{FIM}.
		
		Moreover $\bs{\Delta}_{\bs{\phi}}$ satisfies the following two properties:
		\begin{itemize}
			\item[C1] Asymptotic differentiability (or asymptotic linearity): for all $\bs{\phi},\mb{h} \in \Omega$
			\be
			\label{asyn_lin}
			\bs{\Delta}_{\bs{\phi}+L^{-1/2}\mb{h}} - \bs{\Delta}_{\bs{\phi}} = - \mb{I}(\bs{\phi})\mb{h} + o_P(1),
			\ee
			\item[C2] Asymptotic normality:
			\be
			\label{Delta_G}
			\bs{\Delta}_{\bs{\phi}} \underset{L \rightarrow \infty}{\sim} \mathcal{N}(\mb{0},\mb{I}(\bs{\phi})), \quad \forall \bs{\phi} \in \Omega.
			\ee
		\end{itemize}
	\end{proposition} 
	\textit{Remark}: The proof of Prop. \ref{Prop_LAN} and extensive in-depth discussion about the LAN property can be found in \cite{LeCam60}, \cite[Ch. 6]{LeCam}, and \cite[Ch. 7.6]{vaart_1998}.
	
	Before moving on, it is important to stress that the LAN property can be defined in much more general settings, e.g. for non-i.i.d. observations and for statistical models that do not admit a FIM or even a score vector. Actually, under the regularity conditions in Prop. \ref{Prop_reg}, the expansion in \eqref{LAN} can be thought as the second-order Taylor approximation of the log-likelihood function \cite[Ch. 7.2]{vaart_1998}. Anyway, as said before, even if they are not the weakest ones, the assumptions made in Prop. \ref{Prop_LAN} are satisfied by many data generating processes in SP applications. In particular, they are met by the Elliptical Symmetric (ES) distributions. Specifically, let us define the parametric model of the Real ES (RES) distributions as:
	\be
	\label{RES_par_model}
	\mathcal{P}_{\bs{\phi}} =  \graffe{  p_X | p_X(\mb{x}|\bs{\phi}) = 2^{-N/2} |\mb{V}_1|^{-1/2}  g_0 \left((\mb{x}_l-\bs{\mu})^\top\mb{V}_1^{-1}(\mb{x}_l-\bs{\mu}) \right);  \bs{\phi} \in \Omega },
	\ee  
	and the parameter vector $\bs{\phi}$ is defined in Eq. (6) of our paper as $\bs{\phi} \triangleq \tonde{\bs{\mu}^\top,\ovecs{\mb{V}_1}^\top}^\top$,
	%    	\be
	%    	\label{def_phi}
	%    	\bs{\phi} \triangleq \tonde{\bs{\mu}^\top,\ovecs{\mb{V}_1}^\top}^\top,
	%    	\ee
	where $\bs{\mu} \in \mathbb{R}^N$ is the location vector and $\mb{V}_1 \in \mathcal{M}_N^\mathbb{R}$ is the shape matrix s.t. $[\mb{V}_1]_{1,1}=1$. The general proof of the fact that the RES model in \eqref{RES_par_model} is regular and satisfies the LAN property in Prop. \ref{Prop_LAN} has been provided by Hallin and Paidaveine in \cite[Prop. 2.1]{Hallin_P_Annals} (see also \cite[Appendix 1]{Hallin_P_Annals}). As mentioned above, this is of great practical importance because, as proved by Le Cam in \cite{LeCam60}, \cite[Ch. 6]{LeCam}, if a parametric model is Local Asymptotic Normal, then asymptotically efficient estimators of the parameter of interest $\bs{\phi}$ can be built using a \virg{one-step linear correction} to any preliminary $\sqrt{L}$-consistent estimator $\hat{\bs{\phi}}^\star$ of the true parameter vector $\bs{\phi}_0$. 
	
	\subsection{Efficient one-step parametric estimators}
	In parametric setting, the standard procedure to derive efficient estimators is given by the Maximum Likelihood theory. Specifically, given a set of i.i.d. data  $\{\mb{x}_l\}_{l=1}^L$, an asymptotically efficient estimate of the true parameter vector $\bs{\phi}_0 \in \Omega \subseteq \mathbb{R}^q$, if it exists, can be obtained as: 
	\be
	\label{ML_est}
	\hat{\bs{\phi}}_{ML} \triangleq \underset{\bs{\phi} \in \Omega}{\mathrm{argmax}} \sum\nolimits_{l=1}^L \ln p_X(\mb{x}_l|\bs{\phi}).
	\ee 
	As every practitioner knows, solving the optimization problem in \eqref{ML_est} may result to be a prohibitive task and, in some cases, $\hat{\bs{\phi}}_{ML}$ may not even exist or may not be unique \cite[Ch. 6]{Lehm98}. So, it would be useful to figure out a different methodology to derive efficient estimates.
	
	Under the regularity conditions stated in Prop. \ref{Prop_reg}, if $\hat{\bs{\phi}}_{ML}$ exists, then it satisfies: 
	\be
	\label{D_ML}
	\left.  \bs{\Delta}_{\bs{\phi}}(\mb{x}_1,\ldots,\mb{x}_M)\right|_{\bs{\phi} =\hat{\bs{\phi}}_{ML}} \equiv \bs{\Delta}_{\hat{\bs{\phi}}_{ML}} =\mb{0},
	\ee
	where $\bs{\Delta}_{\bs{\phi}}$ is the central sequence defined in \eqref{cent_seq}. Eq. \eqref{D_ML} can be thought as a set of $q$ nonlinear equations, then we can define a new estimator $\hat{\bs{\phi}}$ given by the one-step Newton-Raphson approximate solution of \eqref{D_ML} as:
	\be
	\label{NR}
	\hat{\bs{\phi}}=\tilde{\bs{\phi}} - [\mb{J}_{\bs{\Delta}}(\tilde{\bs{\phi}})]^{-1}\bs{\Delta}_{\tilde{\bs{\phi}}},
	\ee 
	where $\tilde{\bs{\phi}}$ is a \virg{good} starting point and $\mb{J}_{\bs{\Delta}}(\tilde{\bs{\phi}})$ indicates the Jacobian matrix of $\bs{\Delta}_{\bs{\phi}}$ evaluated at $\tilde{\bs{\phi}}$. Note that the approximation in \eqref{NR} is valid even if $\hat{\bs{\phi}}_{ML}$ does not exists. In \cite{LeCam60} and \cite[Ch. 6]{LeCam}, Le Cam formalized and generalized this intuitive procedure by providing an asymptotic characterization of the class of efficient one-step estimators. This fundamental result is summarized in the following theorem (see also \cite[Ch. 5.7]{vaart_1998}). 
	\begin{theorem}
		\label{theo_one_step_par}
		Let $\{\mb{x}_l\}_{l=1}^L$ be a set of i.i.d. observations sampled from the \virg{true} pdf $p_0 \in \mathcal{P}_{\bs{\phi}}$ satisfying the LAN property as in Prop. \ref{Prop_LAN}. Let $\hat{\bs{\phi}}^\star$ any preliminary $\sqrt{L}$-consistent estimator of the true parameter vector $\bs{\phi}_0 \in \Omega$. Then, the one-step estimator
		\be
		\label{on_step_par}
		\hat{\bs{\phi}} = \hat{\bs{\phi}}^\star + L^{-1/2}\mb{I}(\hat{\bs{\phi}}^\star)^{-1}\bs{\Delta}_{\hat{\bs{\phi}}^\star},
		\ee
		has the following properties:
		\begin{itemize}
			\item[P1] $\sqrt{L}$-consistency
			\be
			\sqrt{L} \tonde{\hat{\bs{\phi}} - \bs{\phi}_0} = O_{P}(1),
			\ee
			\item[P2] Asymptotic normality and efficiency
			\be
			\sqrt{L}\tonde{\hat{\bs{\phi}} - \bs{\phi}_0} \underset{L \rightarrow \infty}{\sim} \mathcal{N}(\mb{0},\mb{I}(\bs{\phi}_0)^{-1}),
			\ee
			where $\mb{I}(\bs{\phi}_0)^{-1} = \mathrm{CRB}(\bs{\phi}_0)$ is the Cram\'er-Rao Bound.
		\end{itemize}
	\end{theorem}      
	\begin{IEEEproof}
		Let us start by showing that the expression defining the one-step estimator in \eqref{on_step_par} can be derived directly from the Newton-Raphson approximation in \eqref{NR}, using the asymptotic differentiability property C1, given in Eq. \eqref{asyn_lin}, of the central sequence. Specifically, in analogy with the definition of Jacobian matrix, we have that:
		\be
		\label{jac}
		\mb{J}_{\bs{\Delta}}(\bs{\phi}) \equiv -L^{1/2}\mb{I}(\bs{\phi}) + o_P(1), \quad \forall \bs{\phi} \in \Omega.
		\ee
		Finally, substituting \eqref{jac} in \eqref{NR}, and noticing that $\hat{\bs{\phi}}^\star$ is a good starting point since it is, by definition, in the $\sqrt{L}$-neighborhood of $\bs{\phi}_0$, yields the expression \eqref{on_step_par}.
		
		The proof of the $\sqrt{L}$-consistency property P1 of $\hat{\bs{\phi}}$ can be found in \cite[Sec. 2.5, Th. 2]{BKRW}. To prove the property P2, we start from the intermediate result provided in \cite[Sec. 2.3, Th. 1]{BKRW}, that is $\mb{I}(\bs{\phi})^{-1}\bs{\Delta}_{\bs{\phi}} \underset{L \rightarrow \infty}{\sim} \mathcal{N}(\mb{0},\mb{I}(\bs{\phi})^{-1})$. Consequently, using the fact that $\hat{\bs{\phi}}^\star$ is $\sqrt{L}$-consistent, the asymptotic normality and efficiency of $\hat{\bs{\phi}}$ in \eqref{on_step_par} follows form a direct application of the Slutsky's theorem \cite[Lemma 2.8]{vaart_1998}. Note that the same warning raised up for Prop. \ref{Prop_LAN} holds here for Theorem \ref{theo_one_step_par}. In fact, in \cite[Sec. 2.3, Th. 1 and Sec. 2.5, Th. 2]{BKRW} only the Hellinger differentiability is required, while here we need to assume the existence of the gradient (w.r.t. $\bs{\phi} \in \Omega$) of the log-likelihood function.
	\end{IEEEproof}
	
	Since, as shown in \cite[Prop. 2.1]{Hallin_P_Annals}, the RES model in Eq. \eqref{RES_par_model} satisfies the LAN property, Theorem \ref{theo_one_step_par} can be readily applied to derive a one-step efficient estimator of the true parameter vector $\bs{\phi}_0 \triangleq \tonde{\bs{\mu}_0^\top,\ovecs{\mb{V}_{1,0}}^\top}^\top$. The closed form expressions of the score vector $\mb{s}_{\bs{\phi}}$ (and consequently the one of the central sequence $\bs{\Delta}_{\bs{\phi}}$) and of the FIM $\mb{I}(\bs{\phi})$, needed to implement the estimator in Eq. \eqref{on_step_par}, can be directly obtained by the ones already derived in our previous work \cite{For_SCRB}. Moreover, as preliminary $\sqrt{L}$-consistent estimator we may use: 
	\be
	\label{preliminary_estim}
	\hat{\bs{\phi}}^\star \triangleq \tonde{\hat{\bs{\mu}}_{Ty}^\top,\ovecs{\widehat{\mb{V}}_{1,Ty}}^\top}^\top,
	\ee 
	where $\hat{\bs{\mu}}_{Ty}^\top$ and $\widehat{\mb{V}}_{1,Ty}$ are the joint Tyler's estimates of the location vector and of the shape matrix constrained to have $[\widehat{\mb{V}}_{1,Ty}]_{1,1}=1$ \cite{Tyler1}, \cite{joint_robust_M_est}.
	
	The result in Theorem \ref{theo_one_step_par} would be enough to derive original, asymptotically efficient, estimators of the location vector $\bs{\mu}_0$ and of the shape matrix $\mb{V}_{1,0}$ in the classical parametric context. Here however, we want to go one step further towards the semiparametric framework. 
	
	\subsection{One-step, semiparametric estimators}
	A semiparametric model $\mathcal{P}_{\bs{\phi},g}$ is a set of pdfs parameterized by a finite-dimensional parameter vector $\bs{\phi} \in \Omega \subseteq \mathbb{R}^q$ and by a function $g \in \mathcal{G}$ that usually plays the role of an infinite-dimensional nuisance parameter \cite{BKRW, Tsiatis}. As amply discussed in \cite{For_SCRB} and \cite{For_SCRB_complex} the ES distributions are a perfect candidate to be modeled as a semiparametric model, since we generally do not have any \textit{a priori} information on the actual density generator $g_0$ characterizing the specific distribution of the observations. Specifically, the RES semiparametric model can be expressed as:
	\be
	\label{RES_semipar_model}
	\mathcal{P}_{\bs{\phi},g} =  \graffe{  p_X | p_X(\mb{x}|\bs{\phi},g) = 2^{-N/2} |\mb{V}_1|^{-1/2}  g \left((\mb{x}_l-\bs{\mu})^\top\mb{V}_1^{-1}(\mb{x}_l-\bs{\mu}) \right);  \bs{\phi} \in \Omega, g \in \mathcal{G} },
	\ee  
	where, as for the parametric case, $\bs{\phi} \triangleq \tonde{\bs{\mu}^\top,\ovecs{\mb{V}_1}^\top}^\top$ while $\mathcal{G}$ is the set of all the \virg{proper} density generators, i.e.  $\mathcal{G} = \graffe{ g: \mathbb{R}^{+} \rightarrow \mathbb{R}^{+} \left|   \int_{0}^{\infty}t^{N/2-1}g(t)dt < \infty, \int p_Xd\mb{x} = 1 \right. }$\cite{CAMBANIS1981}. 
	
	The question that we are going to address here is the following: \textit{is it possible to generalize the concept of one-step estimators, as formalized in Theorem \ref{theo_one_step_par}, to semiparametric inference problems?} To answer to this important point, let us start by focusing on the main building blocks needed to derive the one-step estimator $\hat{\bs{\phi}}$ given, for the parametric case, in Eq. \eqref{on_step_par}. As already discussed in the dedicated statistical literature (see e.g. \cite{BKRW,Tsiatis,Hallin_Werker}) and in our recent works \cite{For_EUSIPCO,For_SCRB,For_SCRB_complex}, the semiparametric counterpart of the score vector $\mb{s}_{\bs{\phi}}$ is the \textit{efficient} score vector $\bar{\mb{s}}_{\bs{\phi},g_0}$ defined as (see \cite{For_EUSIPCO} and \cite[Th. IV.1]{For_SCRB}):
	\be\label{eff_score_vect1}
	\bar{\mb{s}}_{\bs{\phi},g_0}(\mb{x}) \equiv \bar{\mb{s}}_{\bs{\phi},g_0} \triangleq \mb{s}_{\bs{\phi}} - \Pi(\mb{s}_{\bs{\phi}}|\mathcal{T}_{g_0}),
	\ee
	where $\Pi(\mb{s}_{\bs{\phi}}|\mathcal{T}_{g_0})$ is the orthogonal projection of the score vector $\mb{s}_{\bs{\phi}}$ in \eqref{score_vect1} on the semiparametric nuisance tangent space $\mathcal{T}_{g_0}$ \cite{Tutorial}, \cite[Ch. 25.4]{vaart_1998}. The semiparametric counterpart of the FIM $\mb{I}(\bs{\phi})$ is the \textit{efficient} semiparametric FIM (SFIM) \cite{For_EUSIPCO},\cite[Th. IV.1]{For_SCRB}:
	\be\label{S_E_FIM1}
	\bar{\mb{I}}(\bs{\phi}|g_0) \triangleq E_{\bs{\phi},g_0}\{\bar{\mb{s}}_{\bs{\phi},g_0}(\mb{x})\bar{\mb{s}}_{\bs{\phi},g_0}(\mb{x})^\top\}.
	\ee
	On the same line of Eq. \eqref{cent_seq}, we introduce the \textit{efficient} central sequence $\overline{\bs{\Delta}}_{\bs{\phi},g}$ simply as:
	\be\label{cent_seq_sem1}
	\overline{\bs{\Delta}}_{\bs{\phi},g}(\mb{x}_1,\ldots,\mb{x}_L) \equiv \overline{\bs{\Delta}}_{\bs{\phi},g} \triangleq L^{-1/2}\sum\nolimits_{l=1}^{L}\bar{\mb{s}}_{\bs{\phi},g}(\mb{x}_l), \quad \forall \bs{\phi} \in \Omega,\; g \in \mathcal{G}. 
	\ee
	
	The natural \virg{semiparametric} generalization of the ML estimating equations in Eq. \eqref{D_ML} would be \cite[Ch. 25.8]{vaart_1998}
	\be
	\label{D_SML}
	\left.  \bs{\Delta}_{\bs{\phi},g}(\mb{x}_1,\ldots,\mb{x}_M)\right|_{\bs{\phi} =\hat{\bs{\phi}}_{ML},g=\hat{g}^\star} \equiv \bs{\Delta}_{\hat{\bs{\phi}}_{ML},\hat{g}^\star} =\mb{0}.
	\ee
	It must be readily noted that the critical difference between the ML estimating equation in \eqref{D_ML} and their semiparametric generalization in \eqref{D_SML} is that the latter involve a preliminary $\sqrt{L}$-consistent, \textit{non-parametric}, estimator $\hat{g}^\star$ of the nuisance function $g$. Unfortunately, as discussed in \cite[Ch. 25.8]{vaart_1998} and in \cite[Ch. 7]{BKRW}, it is generally impossible to find an estimator of the infinite-dimensional nuisance $g$ that converge to the true function $g_0$ at the $O_P(L^{-1/2})$ rate characterizing most of the parametric estimators. Roughly speaking, the non-parametric estimation of a function requires much more data then the ones needed to estimate a finite-dimensional parameter. 
	%In \cite[Ch. 25.8]{vaart_1998}, a methodology to deal with the slow convergence rate of non-parametric estimators of $g_0$ and then, to solve Eq. \eqref{D_SML}, is proposed. However, this approach still require the implementation of a non-parametric estimator of the Donsker class that is not trivial in most of the practical applications, and specifically for semiparametric inference in ES distributions.
	
	For the specific problem of the semiparametric shape matrix estimation in RES distributions, in their seminal work \cite{Hallin_Annals_Stat_2}, Hallin, Oja and Paindaveine proposed a different approach that does not involve the non-parametric estimation of $g_0$, still providing \textit{nearly} semiparametric efficient estimator of  $\bs{\phi} \triangleq \tonde{\bs{\mu}^\top,\ovecs{\mb{V}_1}^\top}^\top$. The basic idea developed in \cite{Hallin_Annals_Stat_2} is to split the semiparametric estimation problem at hand in two parts:
	\begin{enumerate}
		\item Assume that the true density generator $g_0$ is known and solve Eq. \eqref{D_SML} to derive a \virg{clairvoyant} semiparametric estimatior $\hat{\bs{\phi}}_{s}$.
		\item Robustify $\hat{\bs{\phi}}_{s}$ by using a distribution-free, rank based, procedure. 
	\end{enumerate} 
	
	To better understand this approach, let us start by analyzing the properties of the clairvoyant efficient central sequence $\bs{\Delta}_{\bs{\phi},g_0}$ of a set of RES distributed data.   
	\begin{proposition}\label{cl_CS}
		Let $\{\mb{x}_l\}_{l=1}^L$ be a set of i.i.d. observations sampled from a RES pdf $p_0 \in \mathcal{P}_{\bs{\phi},g}$ in \eqref{RES_semipar_model}. Then, the clairvoyant efficient central sequence $\bs{\Delta}_{\bs{\phi},g_0}$ satisfies the following two properties:
		\begin{itemize}
			\item[CS1] Asymptotic differentiability (or asymptotic linearity): for all $\bs{\phi},\mb{h} \in \Omega$
			\be
			\label{asyn_lin_sem}
			\overline{\bs{\Delta}}_{{\bs{\phi}+L^{-1/2}\mb{h}},g_0} - \overline{\bs{\Delta}}_{\bs{\phi},g_0} = - \bar{\mb{I}}(\bs{\phi}|g_0)\mb{h} + o_P(1),
			\ee
			\item[CS2] Asymptotic normality
			\be
			\label{Delta_G_sem}
			\overline{\bs{\Delta}}_{\bs{\phi},g_0} \underset{L \rightarrow \infty}{\sim} \mathcal{N}(\mb{0},\bar{\mb{I}}(\bs{\phi}|g_0)), \quad \forall \bs{\phi} \in \Omega.
			\ee
		\end{itemize}
	\end{proposition}
	\textit{Remark:} The proof can be found in \cite[Sec. 3]{Hallin_P_Annals}.
	%\begin{IEEEproof}
	%	The proof che be found in \cite[Sec. 3]{Hallin_P_Annals}.
	%\end{IEEEproof}
	
	The result in Prop. \ref{cl_CS} suggests us that, for the semiparametric RES estimation problem at hand, it may be possible to derive semiparametric and asymptotically efficient estimators using a procedure similar to the one provided in Theorem \ref{theo_one_step_par}, simply by substituting the parametric score vector and FIM with their semiparametric counterparts. This intuition is formalized by the next theorem that is also given in our main paper as Theorem 1.
	\begin{theorem}
		\label{theo_one_step_semipar1}
		Let $\{\mb{x}_l\}_{l=1}^L$ be a set of i.i.d. observations sampled from a RES distribution with pdf $p_0 \in \mathcal{P}_{\bs{\phi},g}$ in \eqref{RES_semipar_model}. Let $\hat{\bs{\phi}}^\star$ be any preliminary $\sqrt{L}$-consistent estimator of the true parameter vector $\bs{\phi}_0\triangleq \tonde{\bs{\mu}_0^\top,\ovecs{\mb{V}_{1,0}}^\top}^\top$. Then, the clairvoyant semiparametric one-step estimator
		\be
		\label{one_step_par_semi1}
		\hat{\bs{\phi}}_s = \hat{\bs{\phi}}^\star + L^{-1/2}\bar{\mb{I}}(\hat{\bs{\phi}}^\star|g_0)^{-1}\overline{\bs{\Delta}}_{\hat{\bs{\phi}}^\star,g_0},
		\ee
		has the following properties:
		\begin{itemize}
			\item[PS1] $\sqrt{L}$-consistency
			\be
			\sqrt{L} \tonde{\hat{\bs{\phi}}_s - \bs{\phi}_0} = O_{P}(1),
			\ee
			\item[PS2] Asymptotic normality and efficiency
			\be
			\sqrt{L}\tonde{\hat{\bs{\phi}}_s - \bs{\phi}_0} \underset{L \rightarrow \infty}{\sim} \mathcal{N}(\mb{0},\bar{\mb{I}}(\bs{\phi}_0|g_0)^{-1}),
			\ee
			where $\bar{\mb{I}}(\bs{\phi}_0|g_0)^{-1} = \mathrm{CSCRB}(\bs{\phi}_0|g_0)=\mathrm{CSCRB}(\bs{\mu}_0,\mb{V}_{1,0}|g_0)$ and the constrained semiparametric CRB (CSCRB) \cite{For_SCRB} is evaluated for the constraint $[\mb{V}_{1,0}]_{1,1}=1$.
		\end{itemize}
	\end{theorem}
	\begin{IEEEproof}
		The expression of the semiparametric one-step estimator in \eqref{one_step_par_semi1} can be obtained using the same arguments discussed in Theorem \ref{theo_one_step_par}. The proof of the $\sqrt{L}$-consistency property PS1 of $\hat{\bs{\phi}}_s$ can be found in \cite[Sec. 7.8, Th. 1]{BKRW}. To prove the asymptotic normality, we start from the intermediate result, given in \cite[Sec. 3.3, Th. 2]{BKRW}, that $\bar{\mb{I}}(\bs{\phi}|g_0)^{-1}\overline{\bs{\Delta}}_{\bs{\phi},g_0} \underset{L \rightarrow \infty}{\sim} \mathcal{N}(\mb{0},\bar{\mb{I}}(\bs{\phi}|g_0)^{-1})$. Then, from the expression \eqref{one_step_par_semi1} and from the fact that $\hat{\bs{\phi}}^\star$ is $\sqrt{L}$-consistent, the asymptotic normality and efficiency property PS2 of $\hat{\bs{\phi}}_s$ follows from a direct application of the Slutsky's theorem (see also \cite[Sec. 7.8, Cor. 1]{BKRW}). Again, here we need to assume the existence of the gradient (w.r.t. $\bs{\phi} \in \Omega$) of the log-likelihood function, while in the proof \cite[Sec. 7.8, Th. 1]{BKRW} only the Hellinger differentiability is required.
	\end{IEEEproof}   
	
	As previously underlined and as we can see from its closed form expression in \eqref{one_step_par_semi1}, the clairvoyant estimator $\hat{\bs{\phi}}_s$ relies on the true density generator $g_0$, so it is not useful for inference problems in the semiparametric model \eqref{RES_semipar_model} where the density generator is an unknown nuisance function. However, it has the fundamental role to link the parametric one-step Le Cam's estimator in \eqref{on_step_par} with a distributionally robust estimator of the shape matrix, as shown in \cite{Hallin_Annals_Stat_2} and recalled in Section III of our paper.
	
	\section{Numerical analysis for real $t$-distributed data}\label{numerical1}
	This Section mimics Sec. V of the main paper and provides a numerical investigation about the statistical performance of the \textit{real} $R$-estimator in Eq. (38) in \textit{real} $t$-distributed data.
	
	As in the main paper, in order to distinguish different estimators, each of them will be indicated as $\widehat{\mb{V}}_{1,\gamma}^\varphi$ where $\gamma$ and $\varphi$ specify the estimator at hand. Moreover, we re-normalized $\widehat{\mb{V}}_{1,\gamma}^\varphi$ in order to have $\trace{\widehat{\mb{V}}_{1,\gamma}^\varphi} = N$, i.e. $\widehat{\mb{V}}_\gamma^\varphi \triangleq N \widehat{\mb{V}}_{1,\gamma}^\varphi/\trace{\widehat{\mb{V}}_{1,\gamma}^\varphi}$. 
	
	As a reference, in the figures we also report the Constrained Semiparametric CRB (CSCRB) derived, in closed form, in \cite{For_SCRB}. As performance index for the shape matrix estimators, we use
	\be
	\varsigma_\gamma^\varphi \triangleq \norm{E\{\mathrm{vecs}(	\widehat{\mb{V}}_\gamma^\varphi-\mb{V}_0)\mathrm{vecs}(\widehat{\mb{V}}_\gamma^\varphi-\mb{V}_0)^\top\}}_F,
	\ee
	Similarly, as performance bound, we adopt the index:
	%\be
	%\varepsilon_{CCRB} \triangleq \norm{[\mathrm{CCRB}(\bs{\Sigma}_0)]}_F,
	%\ee
	\be
	\varepsilon_{CSCRB} \triangleq \norm{[\mathrm{CSCRB}(\bs{\Sigma}_0,g_0)]}_F.
	\ee
	
	Unlike the main paper, where a set of complex GG-distributed data are considered, here we generate the dataset according to a real $t$-distribution. The density generator for the $t$-distribution is \cite{CAMBANIS1981}: \footnote{Note that the expression of the density generator in \eqref{g_t} can be obtained from the one given in \cite[Eq. (75)]{For_SCRB} by putting $\eta = 1$.}
	\be\label{g_t}
	g_0(t) \triangleq \frac{2^{N/2}\Gamma(\frac{\lambda+N}{2})}{(\lambda\pi)^{N/2}\Gamma({\lambda}/2 )}\left( 1 + \frac{t}{\lambda} \right)^{-\frac{\lambda+N}{2}}, \; t\in \mathbb{R}^+ 
	\ee
	and the degrees of freedom $\lambda \in (0,\infty)$ controls the non-Gaussianity of the data. In particular, for small values of  $\lambda$ the data are highly non-Gaussian while, as $\lambda \rightarrow \infty$, the distribution collapses into the Gaussian one. The simulation parameters for this study case are:
	\begin{itemize}
		\item $[\bs{\Sigma}_0]_{i,j} = \rho^{|i-j|}, \; i,j=1,\ldots,N$; $\rho = 0.8$ and $N=8$.
		\item The \virg{small perturbation} matrix $\mb{H}^0$ is chosen to be a symmetric random matrix s.t. $\mb{H}^0 = (\mb{G}+\mb{G}^T)/2$ where $[\mb{G}]_{i,j} \sim \mathcal{N}(0,\upsilon^2)$, $[\mb{G}]_{1,1}=0$ and $\upsilon = 0.01$. Note that $\upsilon$ should be small enough to guarantee that $\widehat{\mb{V}}_1^\star + L^{-1/2}\mb{H}^0 \in \mathcal{M}_N^\mathbb{R}$.
	\end{itemize}
	
	As discussed in the main paper, the $R$-estimator in Eq. (38) depends on two \virg{user-defined} quantities: 1) the preliminary estimator $\widehat{\mb{V}}^\star_1$ and 2) the score function $K_g$. In order to assess the impact of their choice on the performance of the $R$-estimator, we perform our simulations by using the Tyler's and the Huber's estimators as preliminary estimators. Moreover, for the Huber's estimator, three different values of the tuning parameter $q$ (i.e. $q=0.9, 0.5, 0.1$) has been adopted \cite[Sec. V.C]{Esa}. Moreover, as score functions, we exploit the \textit{van der Waerden} one and the $t_\nu$-score for $\nu = 0.1, 1, 5$, given in Eqs. (34) and (35) of the main paper. As we will see in the following, the simulation results obtained for the real case are perfectly in line with the one reported in the main paper for the complex case.
	
	\subsection{Semiparametric efficiency}\label{sec_sem_eff1}
	In Figs. 1(a) and 1(b), MSE indices of the \text{real} $R$-estimator in Eq. (38) are plotted as function of the number $L$ of  $t$-distributed observations with $\lambda = 5$ and then compared with the CSCRB. Specifically, in Fig. 1(a) the asymptotic efficiency of the $R$-estimator, exploiting a \textit{van der Waerden} score, is investigated for the two considered preliminary estimators, i.e. Tyler's and Huber's one. As for the complex case, the impact of the choice of the preliminary estimator on the efficiency of the $R$-estimator is negligible. Similarly, the asymptotic impact of the choice of the score functions is also negligible, as shown in Fig. 1(b).
	However, as for the complex case, the score function plays a role in the \virg{finite-sample} performance of the estimator. To see this, in Fig. 1(c), we report the MSE indices obtained for the \textit{van der Waerden} and $t_\nu$- scores as function of the degrees of freedom $\lambda$ for $L=5N$. Note that, for $\lambda = 5$, the $t_5$-score is perfectly specified and then it provides the lowest MSE value at $\lambda = 5$. However, as for the complex case, the \textit{van der Waerden} score confirms its surprisingly good performance (see the discussion on the \virg{Chernoff-Savage result} provided in the main paper).
	
	The $t_\nu$-scores are more flexible since the additional parameter $\nu$ can be used to tune the desired trade-off between semiparametric efficiency and robustness to outliers, as we will see ahead. In particular, $t_\nu$-scores characterized by a small value of $\nu$ increases the robustness of the resulting $R$-estimator at the price of a loss of efficiency. On the other hand, larger values of $\nu$ will provide a better efficiency, sacrificing the robustness as addressed in the next section.

	\begin{figure}\label{fig_eff}
		\centering
		\begin{subfigure}[b]{0.4\textwidth}
			\centering
			\includegraphics[width=\textwidth]{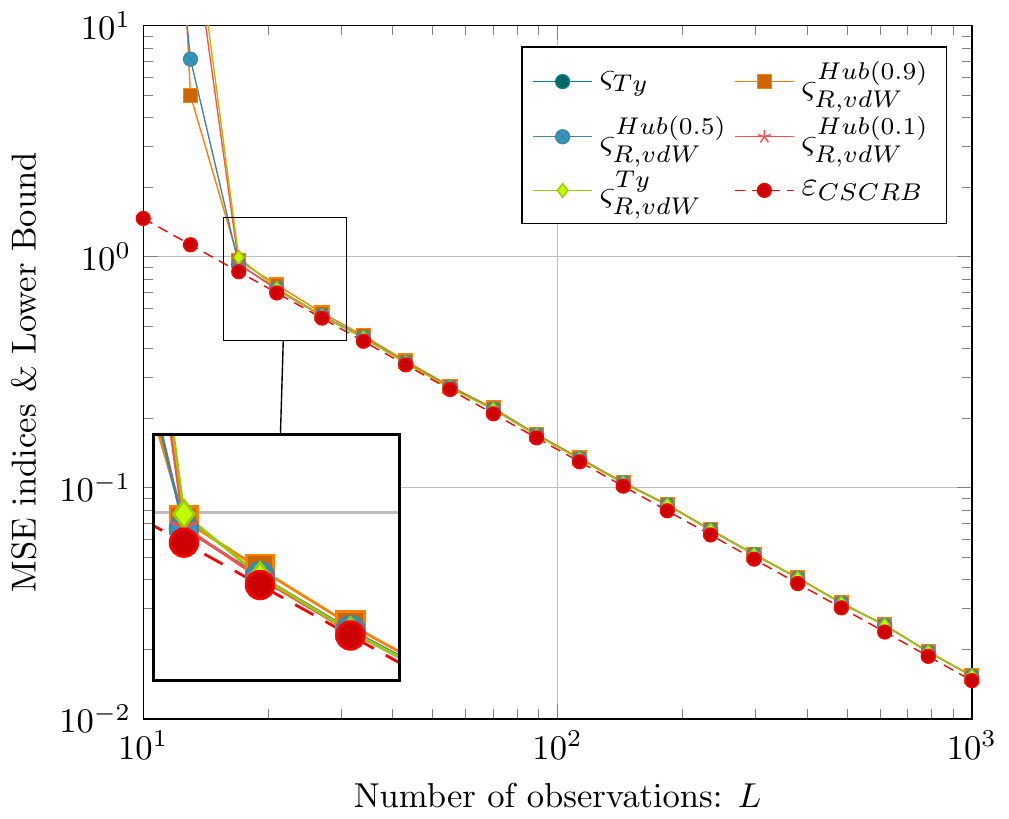}
			\caption{MSE indices vs preliminary Tyler's and Huber's estimators as function of $L$ ($\lambda = 5$).}
		\end{subfigure}
		\begin{subfigure}[b]{0.4\textwidth}
			\centering
			\includegraphics[width=\textwidth]{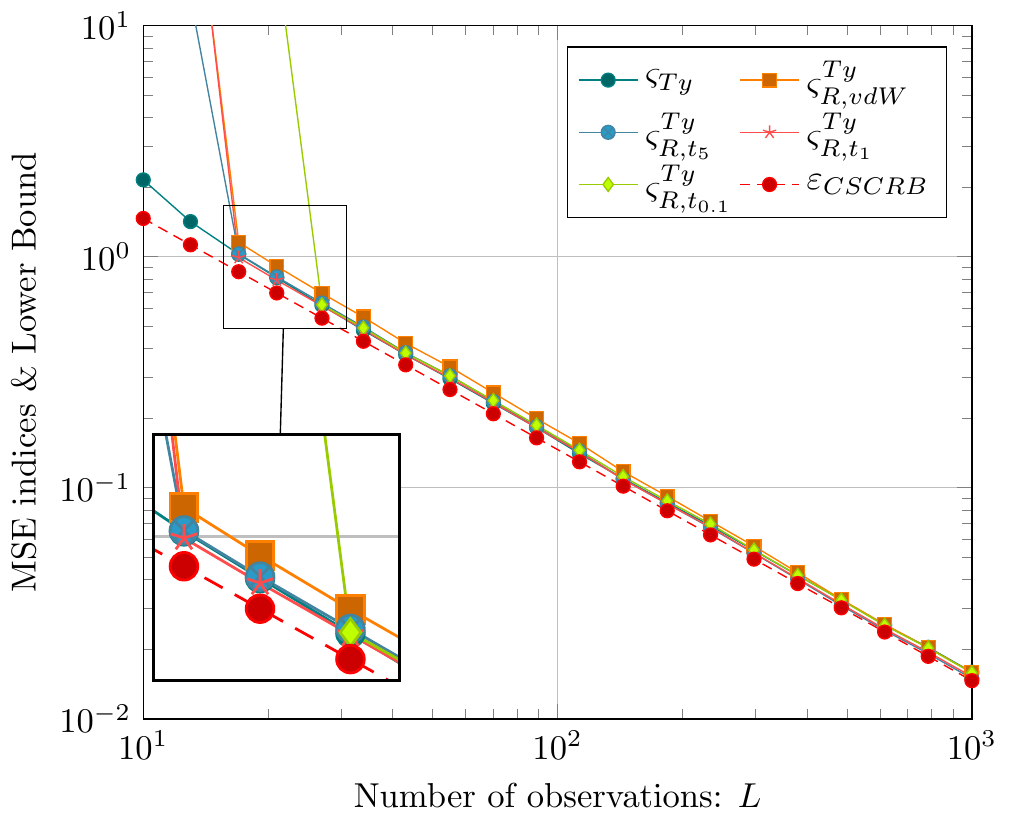}
			\caption{MSE indices vs different score functions $K_g$ as function of $L$ ($\lambda = 5$).}
		\end{subfigure}
		\begin{subfigure}[b]{0.4\textwidth}
			\centering
			\includegraphics[width=\textwidth]{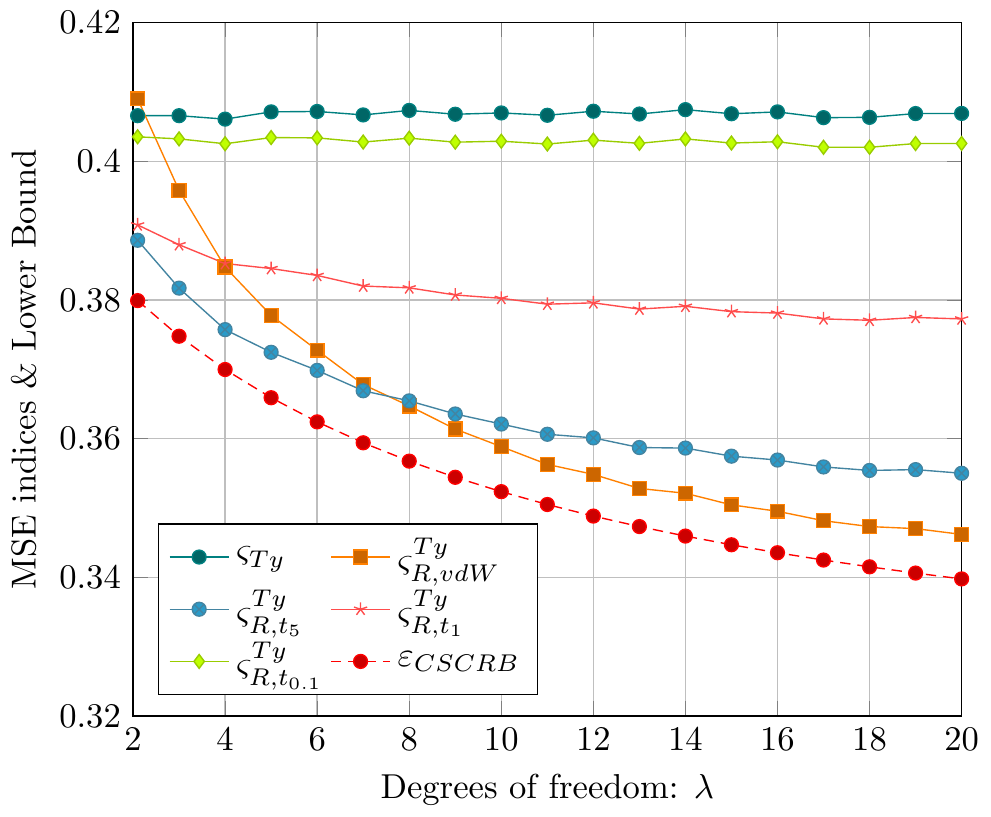}
			\caption{MSE indices vs different score functions $K_g$ as function of $\lambda$ ($L = 5N$).}
		\end{subfigure}
		\caption{MSE performance of the real $R$-estimator.}
	\end{figure}

	\subsection{Robustness to outliers}
	Following Sec. V.B of the main paper, in this subsection we evaluate the \virg{finite-sample} Breakdown Point (BP) \cite{BP_Ty} and the Empirical Influence Function (EIF) \cite{Croux} for the real $R$-estimator in Eq. (38). 
	
	We indicate with $X \triangleq \{\mb{x}_l\}_{l=1}^L \sim RES(\mb{0}, \mb{V}_1,g_0)$ the \virg{pure} $t$-distributed data set whose $g_0$ is given in \eqref{g_t} and with $X_\varepsilon \triangleq \{\mb{x}_{l}\}_{l=1}^L \sim f_{X_\varepsilon}$ the \textit{$\varepsilon$-contaminated} data set s.t.:
	\be\label{cont_model1}
	f_{X_\varepsilon}(\mb{x}|\mb{V}_1,g_0,\varrho) = (1-\varepsilon)  RES(\mb{0}, \mb{V}_1,h_0)+ \varepsilon q_X(\varrho), 
	\ee
	where $\varepsilon \in [0,1/2]$ is a contamination parameter. The function $q_X(\varrho)$ represents the pdf of an outlier $\tilde{\mb{x}}$ that we arbitrary choose to be as $\tilde{\mb{x}} = \tau^{-1} \mb{u}$ where $\mb{u} \sim \mathcal{U}(\mathbb{R}S^{N-1})$ while $\tau \sim \mathrm{Gam}(\varrho,1/\varrho)$ and $\mathrm{Gam}$ indicates the Gamma distribution. The reader can find additional discussion about this model in Sec. V.B of the main paper.

	Let $\widehat{\mb{V}}_\gamma^\varphi(X)$ and $\widehat{\mb{V}}_\gamma^\varphi(X_\varepsilon)$ be two shape matrix estimators evaluated from the pure and the $\varepsilon$-contaminated data sets, respectively. As for the complex case, the finite-sample BP curves can be evaluated as \cite{BP_Ty}:
	\be
	BP_\gamma^\varphi(\varepsilon) \triangleq \mathrm{max} \graffe{\lambda_{\gamma,1}^\varphi(\varepsilon), 1/\lambda_{\gamma,N}^\varphi(\varepsilon) },
	\ee  
	where $\lambda_{\gamma,i}^\varphi(\varepsilon)$ is the $i$-th ordered eigenvalue of the matrix $[\widehat{\mb{V}}_\gamma^\varphi(X)]^{-1}\widehat{\mb{V}}_\gamma^\varphi(Z_\varepsilon)$, s.t. $\lambda_{\gamma,1}^\varphi(\varepsilon) \geq \cdots \geq \lambda_{\gamma,N}^\varphi(\varepsilon)$. Note that $BP_\gamma^\varphi(0) = 1$.
	
	Fig. 2(a) reports the BP curves of the real $R$-estimator in Eq. (38) built upon the \textit{van der Waerden} and three $t_\nu$- scores ($\nu = 0.1, 1, 5$). Since $BP_\gamma^\varphi(\varepsilon)$ depends on $X$ and $X_\varepsilon$, we plot its averaged value over $10^4$ realizations of these data sets. For the sake of comparison, we report also the BP value of Tyler's estimator.  The BP of the non-robust Sample Covariance Matrix (SCM) estimator explodes to $10^{17}$ as soon as $\varepsilon \neq 0$, so we do not include it in the plot. As for the complex case, all the BP curves, related to the $R$-estimator in Eq. (38) are bounded (w.r.t. the one of the non robust SCM) and close to the Tyler's one for every value of $\varepsilon$. 
	%On the other hand, A visual inspection of Fig. \ref{fig:Fig_BP_score} confirms us what already said in Sec. \ref{sec_sem_eff}: $t_\nu$-scores with a small value of $\nu$ lead to more robust estimators. In particular, it can be noted that the BP curves of the $R$-estimator with $t_{0.1}$- and $t_{1}$-score functions coincide with the one of Tyler's estimator. % is lower of the one of the $t_{1}$- and $t_{5}$- based $R$-estimators. %Note that the \textit{van der Waerden} score provide an intermediate BP curve. %Even if this \virg{finite-sample} analysis seems to suggest that the $R$-estimator have a BP value close to the one of an $M$-estimator of shape \cite{BP_Ty}, more rigorous proof and in-depth investigation of this fact is required and they are left to future works. 
	
	Let us now focus on the EIF. Similarly to the complex case discussed in our paper, the EIF can be defined as:
	\be\label{sv1}
	EIF_\gamma^\varphi \triangleq (L+1)\norm{\widehat{\mb{V}}_\gamma^\varphi(X) - \widehat{\mb{V}}_\gamma^\varphi(X,\tilde{\mb{x}})}_F,
	\ee 
	where $\tilde{\mb{x}}$ is an outliers distributed according to the pdf $q_X(\varrho)$ defined in Eq. \eqref{cont_model1}. We refer the reader to the main paper for additional discussion on the definition of the EIF in Eq. \eqref{sv1}. 
	%As Eq. \eqref{sv} suggests, the $EIF_\gamma^\varphi$ gives us a measure of the impact that a \text{single} outlier $\tilde{\mb{z}}$ has on the shape matrix estimator $\widehat{\mb{V}}_\gamma^\varphi$ when it is added to the \virg{pure} data set $Z$. Moreover, if $L$ is sufficiently large, the expression in \eqref{sv} is a good approximation of the theoretical IF \cite{Croux}. For this reason, in our simulation we use $L=1000$. 
	%Since $EIF_\gamma^\varphi$ depends on $X$ and $\tilde{\mb{x}}$, we plot its averaged value over $10^4$ realizations of the data set and the outlier. As for the IF, the most important property that the EIF of a robust estimator should have is the boundeness. In fact, this indicates that the impact of a single outlier on the estimation performance is limited. 
	In Fig. 2(b), we report the EIF of the real $R$-estimator in Eq. (38) built upon the \textit{van der Waerden} and three $t_\nu$- scores ($\nu = 0.1, 1, 5$). As benchmark, the EIF of the Tyler's estimator is adopted since it is known that the relevant IF is continuous and bounded \cite{Esa}. On the other hand, the EIF of the non-robust SCM grows rapidly to $10^4$ as the norm of the outlier $\tilde{\mb{x}}$ increases (i.e. when $\varrho \rightarrow 0$), so we do not include it in the plot. As for the complex case, Fig. 2(b) shows that the EIFs of the $R$-estimator Eq. (38) remain bounded and close to the one of the Tyler's estimator for arbitrary large vale of $\norm{\tilde{\mb{x}}}$ ($\varrho \rightarrow 0$). 
	%Moreover, the results on the robustness induced by the parameter $\nu$ of the $t_\nu$-score are confirmed by the EIF curves: lower values of $\nu$ lead to lower value of the EIF in the presence of an arbitrary large outlier, and consequently, to a more robust $R$-estimator. 
	
	\begin{figure}\label{fig_rob}
		\centering
		\begin{subfigure}[b]{0.4\textwidth}
			\centering
			\includegraphics[width=\textwidth]{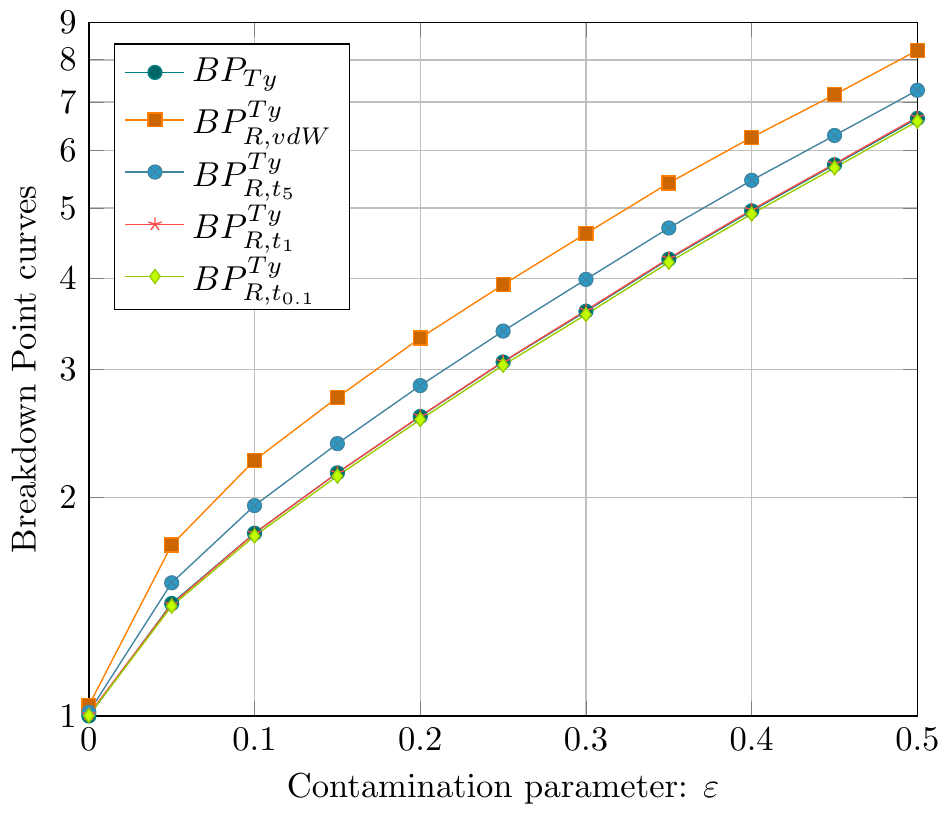}
			\caption{BP vs different score functions $K_g$ as function of $\varepsilon$ ($L=5N$,$\varrho=0.1$,$\lambda = 5$).}
		\end{subfigure}
		\begin{subfigure}[b]{0.4\textwidth}
			\centering
			\includegraphics[width=\textwidth]{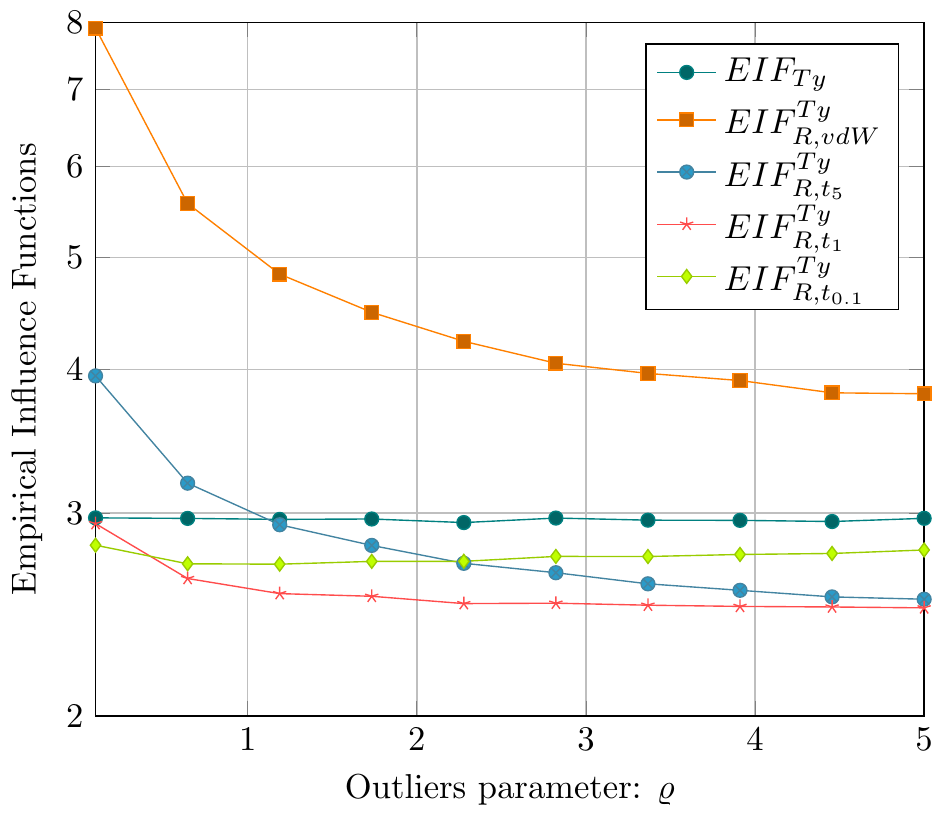}
			\caption{EIF vs different score functions $K_g$ as function of $\varrho$ ($L=1000$,$\lambda = 5$).}
		\end{subfigure}
		\caption{BP and EIF of the real $R$-estimator in $t$-distributed data.}
	\end{figure}

\end{normalsize}

%\bibliographystyle{IEEEtran}
%\bibliography{ref_semipar_eff_estim_sm}

\end{document}